\documentclass[twocolumn,prl,showpacs,amsfonts,amsmath,assymb,eufrak
,longbibliography
]{revtex4-1}
\usepackage{epsfig}
\usepackage{color}
\usepackage{bm}
\usepackage[
colorlinks=true, 
pdfstartview=FitV, 
linkcolor=blue, 
citecolor=magenta, 
urlcolor=blue, 
bookmarks=true,
bookmarksnumbered=true,
pdftitle={},
pdfauthor={}
]{hyperref}
\usepackage{braket}
\usepackage{epsfig,amsopn}
\usepackage{graphicx}
\usepackage{amsmath,amssymb}

\usepackage{calc}

\newcommand{\sectionprl}[1]{{\em #1}\/.---}


\begin{document}

\preprint{APS/123-QED}

\newcommand{\titlename}{Microscopic theory of  fluctuating hydrodynamics in nonlinear lattices}

\preprint{APS/123-QED}

\title{\titlename}% Force line breaks with \\

\author{Keiji Saito$^{1}$, Masaru Hongo$^{2,3}$, Abhishek Dhar$^{4}$, and Shin-ichi Sasa$^{5}$ }
\affiliation{$^{1}$Department of Physics, Keio University, Hiyoshi, Kohoku-ku, Yokohama, Japan}%

\affiliation{$^{2}$Department of Physics, University of Illinois, Chicago, IL 60607, USA}%

\affiliation{$^{3}$RIKEN iTHEMS, RIKEN, Wako 351-0198, Japan}%

\affiliation{$^{4}$International Centre for Theoretical Sciences, Tata Institute of Fundamental Research, Bengaluru, India}%

\affiliation{$^{5}$Department of Physics, Graduate School of Science, Kyoto University, Kyoto, Japan}

\date{\today}%

\begin{abstract}
  The theory of fluctuating hydrodynamics  has been an important tool for analyzing  macroscopic behavior in nonlinear lattices. However, despite its practical success, its microscopic derivation is still incomplete. In this work, we provide the microscopic derivation of  fluctuating hydrodynamics, using the coarse-graining and projection technique; the equivalence of ensembles turns out to be critical. The Green-Kubo (GK) like formula for the bare transport coefficients are presented in a numerically computable form. Our numerical simulations show that the bare transport coefficients exist for a sufficiently large but finite coarse-graining length in the infinite lattice within the framework of the GK like formula. This demonstrates that the bare transport coefficients uniquely exist for each physical system.
\end{abstract}

\maketitle

%%%%%%%%%%

\sectionprl{Introduction}
Hydrodynamics is a universal theory that describes the flow of locally conserved quantities. In addition to the development of numerical computation of complicated flow in macroscopic systems \cite{aw-review}, the concept of hydrodynamics has been extended to nano-fluids \cite{keblinski} and cold atomic systems \cite{doyon,rcdd,ghdexp}, where the standard hydrodynamics in textbooks of fluid dynamics~\cite{landau} cannot be directly applied. In particular, for low dimensional fluids, macroscopic transport coefficients such as heat conductivity diverge due to long-time tail in the correlation functions~\cite{aw,fns,lepri2016book,lepri2003physrep,dhar2008}, which has been experimentally observed in low dimensional materials~\cite{chang1,chang2}. Even for such anomalous transport, it has been recognized that fluctuating hydrodynamics (FH)~\cite{landau,sengers} can provide a quantitative prediction of dynamical phenomena assuming the form of the equations and choice of parameter values~\cite{fns,narayan2002,beijeren2012,spohn2014}.
  A key drawback of the theory is the absence of a microscopic formula for the bare transport coefficients. A naive application of the standard Green-Kubo (GK) formula leads to a divergent answer. The detailed form of the transport coefficients is crucial for our understanding the strong finite-size effects seen in near-integrable models \cite{LLP2020}. In order to deepen our understanding 
it is thus desirable to derive the FH from a microscopic mechanical model and to  connect the parameter values in the hydrodynamic equations with those of the microscopic Hamiltonian.

Let $q_n$ and $p_n$ be variables that represent the position and momentum
of the $n$th particle in a one-dimensional lattice. The Hamiltonian is
generally described as 
\begin{align}
  H&=\sum_{n=1}^{N} {p_n^2 / 2 } + V(r_n ) \, , ~~~~~r_{n}=q_{n+1} - q_n \, ,
  \label{hamil}
\end{align}
where the masses are set to unity and $r_n$ is the stretch variable. The potential $V$ depends solely on the stretch variables. Anomalous heat transport, which refers to  the divergence of the heat conductivity, has been extensively studied for this Hamiltonian~\cite{lepri2016book,lepri2003physrep,dhar2008}. Since there are three locally conserved quantities: the stretch, momentum,
and energy, the long time and large distance behavior of the non-linear
lattice may be described by the effective dynamics of their densities
$ u_{a}(x)$ at position $x$ in the continuous picture, where the subscript $a$ stands for the stretch ($a=r$), momentum ($a=p$) and energy ($a=\epsilon$). According to the FH theory for this system~\cite{beijeren2012,spohn2014}, the time evolution of $ u_{a}(x)$ near equilibrium is
assumed to obey 
\begin{align}
%  \begin{split}
  {\partial_t u_{a} } &=  - \partial_x \bigl[ J_{a,{\rm leq}} (u_{r},u_{p},u_{\epsilon})  \nonumber \\
  &  ~~~~~~~- \sum_{a'=r,p,\epsilon} D_{a,a'}  \partial_x u_{a'}  + \xi_{a ,x} (t)
\bigr] . 
                      \label{nfht-spohn}
\end{align}
Here, $J_{a,{\rm leq}}$ denotes the local equilibrium current which is
given as a function of $(u_r, u_p, u_{\epsilon})$ for each $x$. The functional
form of $J_{a,{\rm leq}}$ is determined from the local equilibrium
thermodynamics or the local equilibrium distribution. The terms $D$ and $\xi$, respectively, stand for dissipation and noise, which are both put by hand in order that the equilibrium properties are guaranteed, imposing the fluctuation dissipation relation~\cite{kth}. Recently, in Ref.~\cite{spohn2014}, Spohn has analyzed the local equilibrium current by transforming the three conserved variables into  left and right moving sound modes, and a heat mode, and consequently derived the nontrivial connection to the Kardar-Parisi-Zhang equation of the nonlinear chains. In addition, through the mode-coupling calculation, the anomalous behavior in the current correlation has been clarified. Later, the scaling form of the space-time correlations arising from hydrodynamics has been numerically confirmed in many types of systems~\cite{das2014,kulkarni15,das19,dksd2019,lzp19,wsbe20}.

% main aim of this Letter

Despite its success, the derivation of FH from Hamiltonian dynamics is still incomplete. In particular, let us focus on the parameter $D_{a,a'}$ which is referred to as the {\it bare transport coefficients} (Below, we use this terminology for all related quantities that are locally transformed).  
These should be distinguished from the macroscopic transport coefficients measured under non-equilibrium conditions, such as heat conductivity. The latter corresponds to renormalized transport coefficients obtained by taking hydrodynamic fluctuations into account. The fundamental problem here is to derive the bare transport coefficients $D_{a,a'}$ from Hamiltonian dynamics. We remark that while the diffusion term formula for integrable chains has been studied in the framework of generalized hydrodynamics~\cite{nbd2018,nbd2019,spohn18}, it is unavailable for non-integrable systems in view of the fact that a simple application gives a divergence in this case. Hence, a more strict and general formulation is necessary to complete the FH theory.

Differences between microscopic expressions of bare transport coefficients and macroscopic transport coefficients have been addressed in the context of projection operator methods \cite{zwanzig,mori_fujisaka,kawasaki,fujisaka}. However, the debate remained formal, and the details on the bare transport coefficients could not be studied due to several uncontrolled functional forms that arise in the derivation. Note that in the mode-coupling calculations in Ref.~\cite{spohn2014}, the assumption of finite bare transport coefficients are critical in deriving diverging heat conductivity. However, the existence of finite bare transport coefficients is still an open question especially in one dimension~\cite{beijeren2012}.
Here we demonstrate that a systematic application of the projection formalism and using ensemble equivalence technique lead to a modification of the standard Green-Kubo formula. This procedure leads to finite bare transport coefficients.
%To fill up this longstanding lacuna, we provide here a systematic derivation of the FH, and derive a GK like formula in a computable form. We numerically obtain a finite value for the bare transport coefficients $D_{a,a'}$ for the Hamiltonian (\ref{hamil}).

\sectionprl{Coarse-graining and projection}
We consider the Hamiltonian (\ref{hamil}) with the total number of sites $N$ and we impose the
periodic boundary conditions $r_{n+N}=r_n$ and $p_{n+N}=p_n$ for the stretch and the momentum variables, respectively~\cite{suppl}. In addition, we introduce the following notations to simply indicate phase-space-dependent conserved quantities at any site $n$:
\begin{align}
\hat{c}_{r,n}:=r_{n} , ~~   \hat{c}_{p,n}:=p_{n} , ~~ \hat{c}_{\epsilon,n}:=p_{n}^2/2+ V(r_n) \, .
\end{align}
Throughout this study, the symbol $\hat{\,}$ on a variable implies that it is a function of the entire phase space $\Gamma$ $(=(r_1,p_1,\cdots , r_N , p_N))$ and hence, the detailed values are given once the phase space is specified. We also denote the current for the conserved quantities $\hat{c}_{a ,n}$ at any site $n$ by $\hat{j}_{a,n}$, which is given as $\hat{j}_{r,n}= - p_n$, $\hat{j}_{p,n}=-\partial V(r_{n-1})/\partial r_{n-1}$, and $\hat{j}_{\epsilon,n}=-p_n \, \partial V(r_{n-1})/\partial r_{n-1}$.

%~\cite{ft0}.

\begin{figure}[t]
  \centering
  \includegraphics[width=80mm]{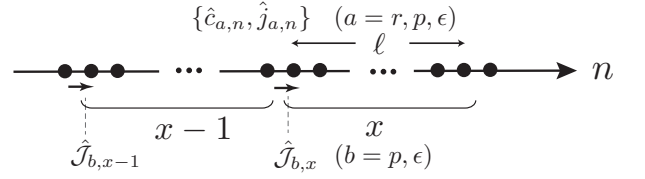} %{./fig1.pdf} 
  \caption{Schematic of the coarse-graining. We define the $x$-coordinate with a unit of $\ell$ sites. The coarse-grained currents $\hat{\cal J}_{p , x}$ and $\hat{\cal J}_{\epsilon , x}$ are locally defined between the blocks.}
\label{fig1}
\end{figure}

  As a first step to obtain the hydrodynamics, we introduce a coarse-graining for conserved quantities:
\begin{align}
  \begin{split}
    \hat{u}_{r,x} &:= (1/\ell) (q_{{\rm G},x+1} - q_{{\rm G},x} ) \, ,  \\
    \hat{u}_{b,x} &:= (1 /\ell) \!\!\! \sum_{n=(x-1)\ell + 1}^{x \ell} \!\! \hat{c}_{b , n} \, , ~~~ (b=p,\epsilon ) \,  ,
    \end{split}     \\
  \begin{split}
    \hat{\cal J}_{r ,x }  &:= - \hat{u}_{p, x} \, , \\
  \hat{\cal J}_{b ,x } & := \hat{j}_{b, (x-1)\ell +1} \, , ~~~~~~~~~~~~~ (b=p,\epsilon ) \,  , 
\end{split}   
  \label{cg}
\end{align}
where the number $\ell$ is the coarse-graining length and hence, we set the total number of sites $N$ to a multiple of $\ell$, and $x=1,\cdots, N/\ell$.
The variable $q_{{\rm G},x}$ is the position of center of mass for the $x$th coarse-graining block defined as $q_{{\rm G},x}=(1/\ell)\sum_{n=(x-1)\ell+1}^{x\ell} q_n$. Note that the coarse-grained stretch variable $\hat{u}_{r,x}$ is a function of microscopic stretch variables $\hat{c}_{r,n}$ \cite{ft0}.
One can easily check that the coarse-grained variable $\hat{u}$ is again a conserved quantity; i.e., the summation of the variables over $x$ is conserved.
The coarse-grained current denoted by $\hat{\cal J}_{a,x}$ is connected to the variable $\hat{u}_{a,x}$ through the continuity equation, i.e., $\partial_t \hat{u}_{a,x}^t= \{ \hat{u}_{a,x}^t , \hat{H} \}=-\nabla_x  \hat{\cal J}_{a ,x }^t$, where the superscript ${}^{t}$ implies the time-dependence and $\{ ..., ...\}$ is the Poisson bracket and the derivative is defined as $\nabla_x A_x :=(1/\ell)(A_{x+1} - A_{x})$ for an arbitrary function $A_x$. For large $\ell$, the variable $\hat{u}_{a , x}$ becomes a macroscopic variable, while the currents $\hat{\cal J}_{p ,x}$ and $\hat{\cal J}_{\epsilon ,x}$ are microscopic variables defined locally at the boundaries between coarse-graining blocks.
See Fig.\ref{fig1}.

Each macrostate defined by the set  $\hat{u}_{a , x}$ corresponds to a large number of microstates and so the evolution of $\hat{u}_{a,x}$ is not deterministic. The internal degrees of freedom  serve like a heat bath providing dissipation and noise that drives the  ``slow'' hydrodynamic fields. As we now show, the  projection formalism allows us to efficiently derive a Fokker-Planck equation for the fields and from this identify the Langevin equations that gives us the required FH in Eqs.~\eqref{nfht-spohn}. Let $\hat{\rho}_t$ be the full phase space density obeying the standard Liouville equation, $\partial_t \hat{\rho}_t =\{ \hat{H} , \hat{\rho}_t\}=:{\mathbb L} \hat{\rho}_t$. Then, we define the following distribution of the  coarse-grained variables:
\begin{align}
    f_{t} ( u )    &:= \int d\Gamma \, \hat{\rho}_t (\Gamma ) \, 
                     \prod_{a=r,p,\epsilon}\,\prod_{x=1}^{N/\ell} \delta (\hat{u}_{a,x} (\Gamma ) - u_{a , x} )                     
    \,   ,
\end{align}
where the integral is defined over the entire phase space. This is the distribution that the variable $\{ \hat{u}_{a ,x} \}$  takes the c-number value $\{ {u}_{a ,x} \}$.
The evolution of $f_t$ is given by:
\begin{align}
&{\partial_t f_t (u) }
                = \partial_t \int d\Gamma \hat{\rho}_t (\Gamma )\,
\prod_{a ,x} \delta (\hat{u}_{a ,x} (\Gamma ) - u_{a , x} )
                 \nonumber \\
  &= \!\int \! d\Gamma \hat{\rho}_t (\Gamma ) \! \sum_{a' , x'} \nabla_{x'} \hat{\cal J}_{a' ,x'} (\Gamma ) {\delta \over \delta u_{a' ,x'} }
 \!  \prod_{a ,x} \delta (\hat{u}_{a, x} (\Gamma ) - u_{a , x} ) \, . \nonumber
%\label{f-evol}
\end{align}
We now use the crucial idea of defining a projection operator~\cite{zwanzig,grabert}  ${\cal P}$ which projects any function $\hat{A}$ onto the coarse-grained conserved variables as
\begin{align}
     {\cal P} \hat{A}(\Gamma)  &= \!\! \int \!\! d\Gamma ' \, \hat{A} (\Gamma ')  \prod_{a,x} \delta (\hat{u}_{a,x} (\Gamma ')\! - \! \hat{u}_{a , x}(\Gamma)  )  /\hat{\Omega}(\hat{u}) \,  , 
                                      \label{projection}
\end{align}
where
the normalization $\hat{\Omega} (\hat{u})$ is defined as $\hat{\Omega} (\hat{u}) =\!\!  \int \!\!  d\Gamma ' \,   \prod_{a,x} \delta (\hat{u}_{a,x} (\Gamma ' ) - \hat{u}_{a , x} (\Gamma)) $. If different phase-space points give the same value in the coarse-grained variables, projected observables also yield the same value between these phase-space points.
The projection redefines a function in terms of coarse-grained conserved quantities. The projection enables us to write $ \hat{\rho}_t = {\cal P} \hat{\rho}_t + {\cal Q} \hat{\rho}_t$,  where ${\cal Q}=1-{\cal P}$,  which  separate the evolution into slow part following the conserved fields and a fast part from the internal degrees. Then from a straightforward calculation which involves using the Markovian approximation (Sec.III in the supplementary material (SM) \cite{suppl})
  , we obtain the Fokker-Planck equation for the distribution $f_t (u)$ (Eq.~(S.19) in SM \cite{suppl}). Finally using  standard  procedure we find the corresponding Langevin equation (Eqs.~(S.21-S.31) in SM \cite{suppl}):
\begin{align}
  \partial_t u_{a, x} &= -\nabla_x \bigl[ \langle \hat{\cal J}_{a , x} \rangle_{\rm LM}^u                             - \sum_{a'} D_{a,a'}^{({\rm A})}  \nabla_x u_{a',x}  \nonumber \\
                           &~~~~~~ - \sum_{a'}  D_{a,a'}^{({\rm S})}  \nabla_x u_{a' ,x} + \xi_{a , x} (t) 
                             \bigr] \, ,  \label{ld}
\end{align}
where the term $\xi_{a ,x} (t)$ is the noise at time $t$ satisfying the fluctuation dissipation relation $\langle\langle \xi_{a , x} (t) \xi_{a ' , x'} (t')\rangle\rangle =
2K_{a,a } \delta_{a ,a '}\delta_{x,x'} \delta (t-t')$ with the bare transport coefficient given explicitly below in (\ref{btc}). The first line indicates the reversible terms, while the second indicates the irreversible terms consisting of noises and bare transport coefficients. The term $ \langle \hat{\cal J}_{a , x} \rangle_{\rm LM}^u$ is the local equilibrium current in (\ref{nfht-spohn}), which turns out to be given as the average with respect to the local microcanonical ensemble $\hat{\rho}_{\rm LM}$:
\begin{align}
    \hat{\rho}_{\rm LM}   &:=  \prod_{a,x} \delta (\hat{u}_{a,x}  - u_{a , x} )    /\Omega ( u ) \, , \label{lmc}
\end{align}
which is the distribution for the values $\{ u_{a , x} \}$ on the phase space. The denominator is a normalization defined as $\Omega ( u )= \int d\Gamma  \prod_{a, x} \delta (\hat{u}_{a,x}(\Gamma) - u_{a , x} )$.
The bare transport coefficient is expressed in terms of the Green-Kubo (GK) like formula as:
\begin{align}
  \begin{split}
\!\!\!\!\!\!  D_{a,a'}^{(\mathrm{S,A)}}
 \! &=  \sum_{a''} (1/2) (K_{a , a ''} \pm K_{a'' , a }) \Lambda_{a' ,a ''}  , \\
  K_{a ,a '} &= \int_{0}^{\infty}  ds\, C_{a , a '} (s) \, , \\
 C_{a , a '} (s) &=  (\ell / N) \langle  (\sum_x  {\cal Q} \hat{\cal J}_{a,x}) (e^{s \mathbb{L} } \sum_{x'} {\cal Q} \hat{\cal J}_{a ' , x'} )\rangle_{\rm eq} ,
  \end{split}  \label{btc}
\end{align}
where $\langle ... \rangle_{\rm eq}$ is the average over the equilibrium distribution $\hat{\rho}_{\rm eq}=e^{-\sum_n (\hat{c}_{\epsilon , n} + P_0 \hat{c}_{r,n})/T}/Z$ with the normalization factor $Z$, since we assume that the dynamics is near equilibrium. Here, the temperature $T$ and the pressure $P_0$ are determined by a given initial state through the total energy and length. The inverse susceptibility matrix element $\Lambda_{a ,a '}$ is explicitly computable~(Sec.VII.A in SM \cite{suppl}).
We note that ${\cal P} \hat{u}_{p,x} = \hat{u}_{p,x}$, and hence we have $K_{r,a}=K_{a,r}=0$. This property as well as ${\bm \Lambda}$ determines the matrix structure of the diffusion matrix ${\bm D}$ \cite{suppl}.

\sectionprl{Computable expressions from the ensemble equivalence}
Let us consider how to practically compute the local microcanonical average on the local equilibrium current term, $\langle \hat{\cal J}_{a , x}\rangle_{\rm LM}^u$, and the projected current that appears in the GK like formula, ${\cal P} \hat{\cal J}_{a ,x}$. From the expressions of local microcanonical ensemble and the projection, we can exactly find simple expressions for the component $a=r$:
 $   \langle \hat{\cal J}_{r , x}\rangle_{\rm LM}^u   = -u_{p,x}$, and ${\cal P} \hat{\cal J}_{r ,x} = -\hat{u}_{p,x} \, $.
Hence, the main focus here is on $\langle \hat{\cal J}_{b , x}\rangle_{\rm LM}^u$ and ${\cal P} \hat{\cal J}_{b ,x}$ with the components $b=p$ and $\epsilon$. As we outline the underlying physics below, we can expand these terms with respect to coarse-grained quantities and variables:
\begin{align}
  \begin{split}
    \!\!\!\!\!\!   \langle \hat{\cal J}_{b , x}\rangle_{\rm LM}^u \! \sim \! A_{b,a} \delta u_{a , x}\! +\! (1/2) H^b_{a,a'} \delta u_{a , x} \delta u_{a', x} \!+\! \cdots \! ,
    \\
    \!\!\!\!\!\!\! ( {\cal P} \hat{\cal J}_{b , x} )  \!\sim  \! A_{b,a} \delta \hat{\tilde{u}}_{a\! , x} \!+ \!(1/2) H^b_{a,a'} \delta \hat{\tilde{u}}_{a \!, x} \delta \hat{\tilde{u}}_{a'\! , x} \!+\! \cdots \! ,
  \end{split}
  \label{appro}
\end{align}
where $\hat{\tilde{u}}$ is defined as $\hat{\tilde{u}}_{r,x}:=(1/\ell)\sum_{n=(x-1)\ell+1}^{x \ell} r_n$ and $\hat{\tilde{u}}_{b',x}:=\hat{u}_{b',x}$ with $b'=p,\epsilon$. In the above expansions, the same subscripts are summed. The symbol $\delta ...$ implies the deviation from the equilibrium value. The matrix elements in ${\bm A}$ \cite{ft4} and ${\bm H}$ are identical to coefficients in the local equilibrium currents of the nonlinear FH in Ref.~\cite{spohn2014}. 

We now outline the underlying mechanism leading to the above computable expressions. It is convenient to discuss $\langle \hat{\cal J}_{b , x}\rangle_{\rm LM}^u$ first. Note that the coarse-grained currents are defined at local sites in the coarse-graining block with the length $\ell$, as depicted in Fig.\ref{fig1}, while the coarse-grained variable $\hat{u}$ is a hydrodynamic variable for sufficiently large $\ell$. We then employ the standard argument in statistical physics; the microcanonical average can be accurately replaced by the canonical average to calculate local observables, as long as the size is large. Leaving the detailed justification in SM \cite{suppl}, we can use the following {\it ensemble equivalence} for large $\ell$ to describe the zeroth order of the gradient expansion in terms of the hydrodynamic motions 
\begin{align}
\hat{\rho}_{\rm LM}  \cong   \hat{\rho}_{\rm LG}  \, ,
\end{align}
where $\hat{\rho}_{\rm LG}$ is the local Gibbs ensemble defined as 
\begin{align}
  \hat{\rho}_{\rm LG}  &= \prod_{x} \hat{\rho}_{\rm LG}^{(x)} \, , ~~ \hat{\rho}_{\rm LG}^{(x)} =
                         e^{-\sum_{a=r,p,\epsilon} \lambda_{a,x} (t)\, \hat{\tilde{u}}_{a,x}     } / { Z_x}  \, . \label{lg}
\end{align}
Here, $Z_x$ is the normalization, and the parameter $ \lambda_{a ,x}$ is a conjugate parameter to the variable $\hat{\tilde{u}}_{a,x}$ that is determined through the condition $ \langle \hat{\tilde{u}}_{a , x} \rangle_{\rm LG} = u_{a,x}$, where $\langle ...\rangle_{\rm LG}$ is an average with the local Gibbs ensemble. This argument systematically yields the expansion for the local equilibrium current in (\ref{appro}).
Next, we can similarly discuss the projected current that appears in the GK like formula. We note that $\langle \hat{\cal J}_{a , x}\rangle_{\rm LM}^u$ can be obtained in ${\cal P} \hat{\cal J}_{a ,x} $ by replacing a phase-space-dependent variable $\hat{u}_{a,x}$ by a c-number value $u_{a ,x}$ (see definitions (\ref{projection}) and (\ref{lmc})). This indicates that the projected current is accurately computable with the ensemble equivalence technique as above, which leads to the expansion in (\ref{appro}) \cite{suppl}.

\begin{figure}[t]
\centering
\includegraphics[width=80mm]{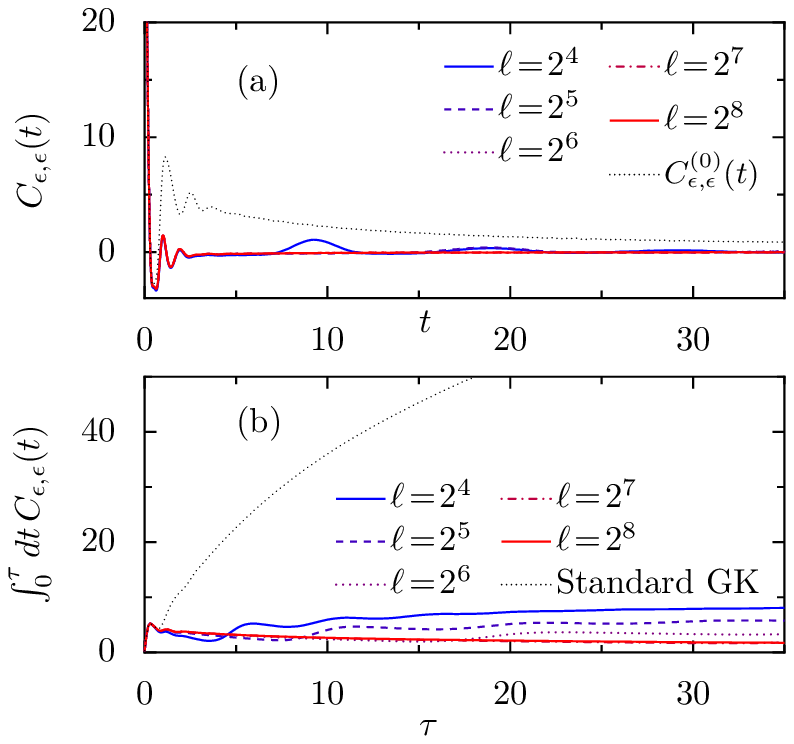}%{./fig2.pdf}
\caption{Numerical demonstration of the GK like formula (\ref{btc}) for the element $(\epsilon,\epsilon)$. Parameters: $k_3=2.0,k_4=1.0, T=3.0$ and $N=2^{15}$. (a): The correlations as a function of time for different $\ell$. $C_{\epsilon , \epsilon}^{(0)}$ is the standard energy current correlation for obtaining the macroscopic heat conductivity. (b): Integration of the correlations up to $\tau$. `Standard GK' (black dotted line) implies $\int_{0}^{\tau} dt \, C_{\epsilon , \epsilon}^{(0)} (t)$, which shows clear divergence. The integration for finite $\ell$ shows the convergence, where the saturated values are plotted in Fig.\ref{fig3}.}
\label{fig2}
\end{figure}

\sectionprl{Numerical investigation}
In the remainder of this paper, we perform a numerical calculation in order to see how unique bare transport coefficients emerge. We use the Fermi-Pasta-Ulam-Tsingou (FPUT) chain with the potential term:
\begin{align}
  V(r) &= (1/2)\, r^2  + (k_3/3) \, r^3  + (k_4 /4)\, r^4 \, .
\end{align}
We remark that the hydrodynamics behavior has been numerically checked in this model~\cite{das2014}.

We show the typical behavior of the correlation function $C_{a ,a '} (t)$. Here, we present the most important element, the energy-energy current correlation function $C_{\epsilon , \epsilon} (t)$ because the standard energy current correlation for obtaining macroscopic transport coefficient shows a power-law decay at long times, resulting in a diverging heat conductivity. 
We present the other elements in SM~\cite{suppl}. In Fig.\ref{fig2}(a), we show the time-dependence of $C_{\epsilon , \epsilon} (t)$ for many values of $\ell$ for the system size $N=2^{15}$ and temperature $T=3.0$ without pressure; the system parameters are $(k_3, k_4)=(2.0,1.0)$~\cite{ft6}. For small $\ell$, we observe small humps in the time-domain. These humps occur every $\ell/c$ where $c$ is the sound velocity $(c\sim 1.54)$ reflected from the sound propagation~\cite{ft7}. As $\ell$ increases, the amplitudes of humps decrease and the overall functional structures collapses onto the same curve, where finite values are seen only at the small-time scale. For comparison, we also show the standard energy current correlation denoted by $C_{\epsilon , \epsilon}^{(0)}(t)$, which corresponds to $C_{\epsilon , \epsilon} (t)$ with $\ell =1$ where the projection contains only the first order dropping the higher orders. In Fig.\ref{fig2}(b), integration up to $\tau$ is shown for the correlation functions in Fig.\ref{fig2}(a). The integration of standard energy current correlation denoted by `Standard GK' is also presented, which shows a clear divergence. In contrast, the integral of $C_{\epsilon , \epsilon} (t)$ with finite $\ell$ converges for sufficiently large $\ell$. The main contribution in the saturated integration is given from the short-time behavior in the correlation. 
\begin{figure}[t]
\centering
\includegraphics[width=80mm]{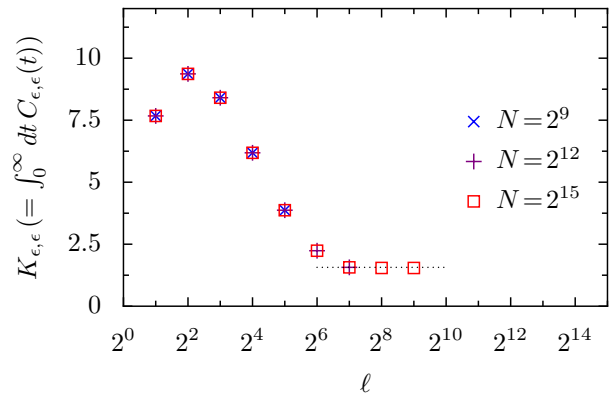}%{./fig3.pdf}
\caption{Bare transport coefficients versus coarse-graining length $\ell$. For same system parameters in Fig.\ref{fig2}, we computed the integration of the GK like formula with different $\ell$ for three system sizes: $N=2^9, 2^{12}$ and $2^{15}$. The values for the same $\ell$ do not differ between different $N$, and eventually saturate for sufficiently large $\ell$. This implies that the bare transport coefficients for each system can be uniquely determined for $1 \ll \ell \ll N$. The integration are performed up to $\tau=100$ as in Fig.\ref{fig2}(b) for all cases.}
\label{fig3}
\end{figure}

In Fig.\ref{fig3}, we show the bare transport coefficients $K_{\epsilon,\epsilon}$ computed via the GK like formula for different coarse-graining lengths $\ell$. Particularly, we consider three different system sizes, $N=2^9, 2^{12}$ and $2^{15}$, and compute the bare transport coefficients for different $\ell$. The figure shows that the same coarse-graining length give same values even when the system sizes are different. For sufficiently large coarse-graining length, the bare transport coefficients are uniquely determined. We stress that the order of limitation in the formula (\ref{btc}) is critical, i.e., $  K_{a, a '} =\lim_{\tau\to\infty}\lim_{N \to \infty} \int_0^{\tau} ds \, C_{a , a '} (s ) $ with the condition $1 \ll \ell \ll  N$. Using the saturated functional form for sufficiently large $\ell$, one can estimate the values of bare transport coefficients $(K_{p,p}, K_{p,\epsilon} (=\!\!-K_{\epsilon , p} ), K_{\epsilon , \epsilon})\sim (0.2\times 10 , 0.2\times 10^{-2}, 0.1\times 10)$.

\sectionprl{Summary}
To summarize, we presented a microscopic theory to derive fluctuating hydrodynamics (FH) in nonlinear lattices. The formalism presented here is quite general and it would be an interesting problem to compare the results of the lattice system to fluid systems that have been studied so far~\cite{miron, zubarevmorozof, sasa2014, hongo,medenjak}.
We hope that the microscopic theory presented here can provide a resolution of some of the open issues in low-dimensional transport~\cite{LLP2020} and useful information for other applications~\cite{sasa1}, and also gives a possibility to extend the FH to other classes of many-body systems such as~\cite{basile2006prl,tamaki-saito1,tamaki-saito2}.

\bigskip

\sectionprl{Acknowledgement}
K.S. was supported by Grants-in-Aid for Scientific Research (JP16H02211, JP19H05603, JP19H05791).
M.H. was supported by the U.S. Department of Energy, Office of Science, Office of Nuclear Physics under Award Number DE-FG0201ER41195, and the RIKEN iTHEMS Program (in particular iTHEMS STAMP working group).
A.D. acknowledges support of the Department of Atomic Energy, Government of India, under project no.12-R\& D-TFR-5.10-1100. S.S. was supported by KAKENHI (JP17H01148, JP19H05496, JP19H05795). We are grateful to C. Bernardin, B. M. Itami, H. Nakano, M. Sasada, and H. Spohn for their valuable comments.

\bibliography{FPUT.bib}

\begin{thebibliography}{48}
\bibitem [{ft0()}]{ft0}%
  \BibitemOpen
\bibfield  {journal} {  }\href@noop {} {}Namely, one can write the
  coarse-grained stretch variable: $\hat{u}_{r,x}=(1/\ell)\! \sum_{n=(x-1)\ell
  +1}^{x \ell} \sum_{k=1}^{\ell} r_{n + k -1}/\ell $.\BibitemShut {Stop}%
\bibitem [{ft4()}]{ft4}%
  \BibitemOpen
  \href@noop {} {}The first order term in ${\cal P}{\cal J}_{a ,x}$ is reduced
  to the Mori projection in \cite{mori}\BibitemShut {NoStop}%
\bibitem [{ft6()}]{ft6}%
  \BibitemOpen
  \href@noop {} {}We use the velocity-Verlet algorithm with the finite time
  discretization value $dt = 0.01$ and the average is taken over sample numbers
  $10^5$.\BibitemShut {Stop}%
\end{thebibliography}

\clearpage

\pagestyle{empty}

\begin{widetext}
  
%%%%%%%%%%%%%%%%%%%%%%%%%%%%%%%%%%%%%%%%%%%
%%%%%%%%%%%%%%%%%%%%%%%%%%%%%%%%%%%%%%
% To modify figure captions
\makeatletter
\long\def\@makecaption#1#2{{
\advance\leftskip1cm
\advance\rightskip1cm
\vskip\abovecaptionskip
\sbox\@tempboxa{#1: #2}%
\ifdim \wd\@tempboxa >\hsize
 #1: #2\par
\else
\global \@minipagefalse
\hb@xt@\hsize{\hfil\box\@tempboxa\hfil}%
\fi
\vskip\belowcaptionskip}}
\makeatother
%%%%%%%%%%%%%%%%%%%%%%%%%%%%%%%%%%%%%%

%\setcounter{figure}{0}
%\def\thefigure{S.\arabic{figure}}
\setcounter{equation}{0}
\def\theequation{A.\arabic{equation}}
%%%%%%%%%%%%%%%%%%%%%%%%%%%%%%%%%%%%%%%

 \begin{center}
{\large \bf Supplemental Material for \protect \\
  ``Microscopic theory of the fluctuating hydrodynamics in nonlinear lattices'' }\\
\vspace*{0.3cm}
Keiji Saito$^{1}$, Masaru Hongo$^{2,3}$, Abhishek Dhar$^{4}$, and Shin-ichi Sasa$^{4}$
\\
\vspace*{0.1cm}

$^{1}${\small \it Department of Physics, Keio University, Hiyoshi, Kohoku-ku, Yokohama 223-8522, Japan} \\

$^{2}${\small \it Department of Physics, University of Illinois, Chicago, IL 60607, USA}\\

$^{3}${\small \it RIKEN iTHEMS, RIKEN, Wako 351-0198, Japan}\\

$^{4}${\small \it Department of Physics, Graduate School of Science, Kyoto University, Kyoto, Japan}
\end{center}

\setcounter{equation}{0}
\renewcommand{\theequation}{S.\arabic{equation}}
\renewcommand{\thefigure}{S\arabic{figure}}
\renewcommand{\bibnumfmt}[1]{[S#1]}

%\tableofcontents

\section{Setup}
We consider an $N$-particle system. Let $\Gamma$ be the entire phase space; i.e., $\Gamma=(r_1,p_1, r_2,p_2, \cdots , r_N , p_N)$, where ${r}_n$ and ${p}_n$ respectively stand for the values of the stretch and the momentum at any site $n$.
For an arbitrary physical quantity $a$, we use the notation $\hat{a}$ to mean that it is a phase-space dependent variable; i.e., its value is determined as soon as the phase space is specified. For instance, $\hat{r}_n =r_n$ and $\hat{p}_n=p_n$ for the phase space $\Gamma=(r_1,p_1, r_2,p_2, \cdots , r_N , p_N)$. The quantity without the symbol $\hat{~~}$ is a c-number value.  With this notation, the Hamiltonian is described as
\begin{align}
\hat{H} &= \sum_{n=1}^{N} {\hat{p}_n^2 \over 2 } + V (\hat{r}_{n} ) \, , ~~~~~(r_n =q_{n+1} - q_n) \, , \label{hamil:suppl}
\end{align}
with the periodic boundary conditions, $r_{n+N}= r_{n}$ and $p_{n+N}=p_n$. Note that we use the stretch variable $r_n$ instead of the position $q_n$ for the phase space. To obtain the boundary condition $r_{n+N}=r_n$, we prepare an infinite line where infinite number of particles are set. Then, we set the configuration in order that the boundary condition is satisfied. Schematic on the periodic boundary condition is depicted in Fig.\ref{fig1:suppl}(a).

Clearly, we have three conserved quantities: the stretch $(\hat{r}_n$, momentum $\hat{p}_n$, and energy $\hat{\epsilon}_n)$ and the local energy is defined as
\begin{align}
\hat{\epsilon}_n &= {\hat{p}_n^2 \over 2 } + V ( \hat{r}_{n} ) \, .
\end{align}
We use the simple notation $\hat{c}_{a ,n}~(a=r,p,\epsilon)$ to express $\hat{c}_{a ,n}\bigr|_{a=r} = \hat{r}_n$, $\hat{c}_{a ,n}\bigr|_{a=p} = \hat{p}_n$, and $\hat{c}_{a ,n}\bigr|_{a=\epsilon} = \hat{\epsilon}_n$. If we write ${c}_{a ,n}$, it implies a c-number value.

Let $\hat{\rho}_t$ be a density distribution at any time t whose time evolution is determined by the Liouville equation:
\begin{align}
  \partial_t {\hat \rho}_t &= \left\{ \hat{H} , {\hat \rho}_t \right\} \nonumber \\
  &= \sum_{n}
                             (\partial \hat{H}/\partial \hat{q}_n
                             )(\partial \hat{\rho}_t/\partial \hat{p}_n ) -  (\partial \hat{H}/\partial \hat{p}_n )(\partial \hat{\rho}_t/\partial \hat{q}_n )                             \nonumber \\
&= \sum_{n=1}^{N}
                             (\partial \hat{H}/\partial \hat{r}_{n-1} - \partial \hat{H}/\partial \hat{r}_n
  )(\partial \hat{\rho}_t/\partial \hat{p}_n ) -  (\partial \hat{H}/\partial \hat{p}_n )
  (\partial \hat{\rho}_t/\partial \hat{r}_{n-1} - \partial \hat{\rho}_t/\partial \hat{r}_n
  )                                                    =:{\mathbb L}                             {\hat \rho}_t          \, , \label{liov:suppl}
\end{align}
where at the last line, we symbolically write the equation introducing the Liouville operator ${\mathbb L}$. With the Liouville operator, one can write $\hat{\rho}_t =e^{{\mathbb L} t}\hat{\rho}$ where $\hat{\rho}$ is a function of the initial phase space. Similarly, the variable that evolves in time denoted by $\hat{c}_{a , n}^t$ is expressed as a function of the initial phase space as $\hat{c}_{a , n}^t = e^{{\mathbb L}^{\dagger} t} \hat{c}_{a , n}$. 
Furthermore, the continuity equation in this notation is given as
\begin{align}
  \partial_t  \hat{c}_{a , n}^t  & = \left\{ \hat{c}_{a , n}^t , \hat{H} \right\} =
                                       - ( \hat{j}_{a , n+1}^t - \hat{j}_{a , n}^t ) \, ,
\end{align}
through which one finds the expressions for the local currents as follows:
\begin{eqnarray}
  \hat{j}_{r,n} = - {\hat{p}_n } \, , ~~~~
  \hat{j}_{p,n} = -{\partial V (\hat{r}_{n-1}) \over \partial \hat{r}_{n-1} } \, , ~~~~
\hat{j}_{\epsilon ,n} = -{\hat{p}_n } {\partial V(\hat{r}_{n-1}) \over \partial \hat{r}_{n-1}} \, . \label{microcurrent3:suppl}
\end{eqnarray}
\begin{figure}[t]
  \centering
  \includegraphics[width=100mm]{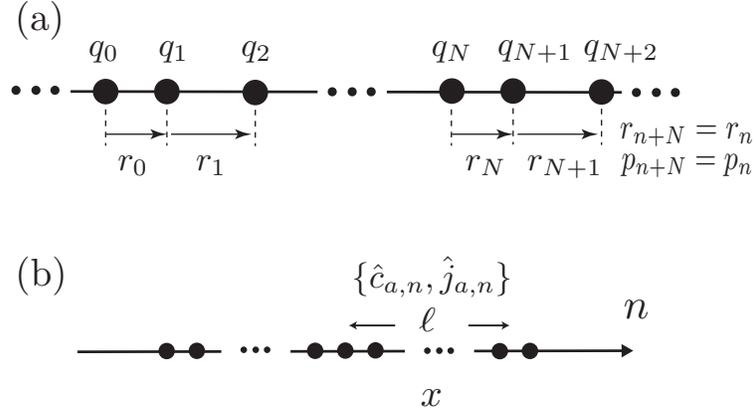} %{./fig1suppl.pdf} 
  \caption{(a): Schematic for the periodic boundary conditions $r_{n+N}=r_n$ and $p_{n+N}=p_n$. With these boundary conditions, the phase space is safely spanned with the variables $\{ r_n, p_n\}$ instead of $\{q_n , p_n \}$. (b): Schematic picture of the coarse-graining.}
  \label{fig1:suppl}
\end{figure}

\section{Coarse-graining}
We introduce the new $x$-coordinate with the unit consisting of $\ell$ sites. We define a coarse-graining on the variables as 
\begin{align}
  \begin{split}
    \hat{u}_{r,x} &:= (1/\ell ) (\hat{q}_{G,x+1} - \hat{q}_{G,x})
    = (1/\ell)\!\!\! \sum_{n=(x-1)\ell +1}^{x \ell} \sum_{k=1}^{\ell} r_{n + k -1}/\ell \, ,   \\
  \hat{u}_{b ,x} &:= (1 /\ell) \sum_{n=(x-1)\ell + 1}^{x \ell} \!\! \hat{c}_{b , n} \, , ~~~b=p,\epsilon ,
\end{split}
                   \label{cg:suppl}
\end{align}
where $x=1,\cdots, N/\ell $, and $\hat{q}_{G,x}$ is a position of the center of mass for the $x$th block, i.e., $\hat{q}_{G,x} =(1/\ell)\sum_{n=1}^{\ell} \hat{q}_{(x-1)\ell + n}$. Note that $\sum_x \hat{u}_{a ,x}$ are conserved for each $a\,(=r,p,\,{\rm and}\,\epsilon)$. Hence, we can define the local current for each coarse-grained variable. The continuity equation for coarse-grained variables $\hat{u}_{a ,x}~(a=r,p,\epsilon)$ is given by
\begin{align}
  \partial_t \hat{u}_{a ,x}^t &=  \left\{ \hat{u}_{a , x}^t , \hat{H} \right\} = - \nabla_x  \hat{\cal J}_{a ,x }^t \, , \label{contmacc:suppl}
\end{align}
where the derivative $\nabla_x$, acting on an arbitrary operator $\hat{A}_x$, is defined as $\nabla_x \hat{A}_x := (1/\ell) (\hat{A}_{x+1} - \hat{A}_x ) $, and the current is given by  
\begin{align}
  \begin{split}
        \hat{\cal J}_{r ,x } &= -\hat{u}_{p, x} \, , \\
    \hat{\cal J}_{b ,x } &= \hat{j}_{b, (x-1)\ell +1} \, ,  ~~~b=p,\epsilon .    \label{currnetcg:suppl}
    \end{split}
\end{align}
The current of coarse-grained stretch variable is directly connected to the coarse-grained momentum, which indicates that the coarse-grained stretch variable
does not have a diffusion term. The current for the momentum and energy term is defined at the edge of the unit sector. 
For a sufficiently large coarse-graining length $\ell$, the coarse-grained variable $\hat{u}_{a , x}$ becomes a macroscopic variable, while the local current $\hat{\cal J}_{b , x} ~(b=p,\epsilon)$ is a microscopic local variable defined at local sites.

\section{The distribution of coarse-graining quantities and the master equation}
We define
\begin{align}
  \delta (\hat{u} - u )  &:= \prod_{a =r,p,\epsilon}\prod_{x=1}^{N/\ell} \delta (\hat{u}_{a, x} - u_{a, x}) \, .
\end{align}
We then define the distribution function of coarse-grained conserved quantities as
\begin{align}
  f_t ( u ) &:= \int d \Gamma \hat{\rho}_t (\Gamma ) \delta (\hat{u} (\Gamma ) - u ) \, .
\end{align}
Here $\int d\Gamma $ denotes the integration over the phase space; i.e., $\int d\Gamma\, \hat{A} (\Gamma) = \prod_{n=1}^{N}\int_{-\infty}^{\infty} dr_n \int_{-\infty}^{\infty} d p_n \, \hat{A}(r_1,p_1, \cdots, r_N , p_N) $ by explicitly writing  the phase-space dependence for the variable. In what follows, we explicitly write the phase-space dependence if necessary.
In addition, we introduce the projection operator ${\cal P}$, which acts on an arbitrary variable $\hat{A}$, as
\begin{align}
  \begin{split}
  ({\cal P} \hat{A}) \left[ \Gamma \right] &:= \int d\Gamma ' \, \hat{A}(\Gamma' ) \, { \delta (\hat{u} (\Gamma ' )  - \hat{u}(\Gamma) )  /  \Omega (  \hat{u}(\Gamma) ) } \, ,  \\
  \Omega (  \hat{u}(\Gamma) )   &:= \int d\Gamma ' \, \delta ( \hat{u} (\Gamma ' )  - \hat{u} (\Gamma )  )  \, .
  \end{split}
\end{align}
Note that if $\hat{u}_{a ,x} (\Gamma_1 ) = \hat{u}_{a ,x} (\Gamma_2 )$ $\forall\,a , \,x$, then  $ ({\cal P} \hat{A}) \left[ \Gamma_1 \right]  =({\cal P} \hat{A}) \left[ \Gamma_2 \right] $. Thus, the projection redefines the observables in terms of conserved quantities. 

We note the following relation:
\begin{align}
{\partial_t f_t (u) }
&= \partial_t \int d\Gamma \hat{\rho}_t (\Gamma )\, \delta ( \hat{u}(\Gamma ) - u ) 
= \int d\Gamma \hat{\rho}_t (\Gamma ) \sum_{a , x} \nabla_x \hat{\cal J}_{a ,x} (\Gamma ) {\delta \over \delta u_{a ,x} } \delta  ( \hat{u}(\Gamma ) - u ) \, .
\end{align}
We insert the formal relation $ \rho_t = {\cal P} \rho_t + {\cal Q} \rho_t$ to get the closed form in terms of the distribution $f_t (u)$.
To this end, we note the expression
\begin{align}
{\cal Q} \hat{\rho}_t & =\int_{-\infty}^t ds e^{(t-s){\cal Q} {\mathbb  L} } {\cal Q}{\mathbb L} {\cal P} \rho_s  
=\int_{-\infty}^{t} ds \int {\cal D} u' e^{(t-s){\cal Q} {\mathbb L} } {\cal Q} \sum_{a', x'} (\nabla_{x'} \hat{\cal J}_{a', x'}) 
\delta (\hat {u} - u' ) {\delta \over \delta u_{a' , x' }'} (f_s (u') / \Omega (u' ))  \, , 
\end{align}
where $\int {\cal D} u := \prod_{a , x}\int d u_{a , x}$.
%where we used the relations $\partial_t \rho_t = {\mathbb L}\rho_t$ and $\partial_t \Gamma_t = {\cal L}^{\dagger} \Gamma_t = - {\cal L} \Gamma_t$.
After some manipulations, one gets the following expression
\begin{align}
  \begin{split}
  {\partial_t f_t (u) } &=  \sum_{a , x} {\delta \over \delta u_{a , x}} ( \langle \nabla_x \hat{\cal J}_{a , x}\rangle_{{\rm LM}}^u  f_t (u) )
\\
&+ \sum_{a , x}\sum_{a ', x'}  {\delta \over \delta u_{a , x}} \int_{-\infty}^t ds \int {\cal D} u' \Omega (u)
(\nabla_x \nabla_{x'} K_{a x,a ' x'} (u,u' ;t-s) ) {\delta \over \delta u_{a ' , x'}}   (f_s (u') / \Omega (u' ))  \, ,\\
K_{a x,a ' x'}(u,u' ; t-s) &= \langle ({\cal Q} \hat{\cal J}_{a ,x}) (e^{(t-s) {\cal Q }{\mathbb L} }  \delta (\hat{u} - u') {\cal Q} \hat{\cal J}_{a ' , x'}
) \rangle_{\rm LM}^u \, , \label{masterftu:suppl}
\end{split}
\end{align}
where $\langle ... \rangle_{\rm LM}^u$ implies the average over a local microcanonical ensemble $\hat{\rho}_{{\rm LM}} (\Gamma )$ defined as 
\begin{align}
  \begin{split}
    \hat{\rho}_{{\rm LM}} (\Gamma )
    &:=\delta (\hat{u}(\Gamma) -u )/\Omega (u) \,  , \\
  \Omega ( u )   &:= \int d\Gamma  \, \delta ( \hat{u} (\Gamma  )  - u  )  \, . \label{phase-suppl}
  \end{split}
\end{align}
This is the microcanonical ensemble for each sector $x$ assigning the c-number values $\{ u_{a ,x}\}$.

We {\it physically} consider the term $K_{a x,a ' x' }(u,u' ; t-s)$ in (\ref{masterftu:suppl}), which is eventually reduced to the bare transport coefficients. Note that $K_{r x, a' x'} = K_{a x, r x'}=0$ due to ${\cal Q}{\cal J}_{r,x} =0$. Hence the main focus below is on the other terms.   
Note that the current $\hat{\cal J}_{b ,x}~(b=p,\epsilon)$ is defined locally at one sector of the $x$-coordinate, as in (\ref{currnetcg:suppl}); see the schematic picture in Fig.\ref{fig2:suppl}. Namely, the term $K_{b x,b ' x' }(u,u' ; t-s)~(b,b'=p,\epsilon)$ is the correlation function between the locally defined observables. We set a sufficiently large coarse-graining length $\ell$ by which the coarse-grained variable $\hat{u}$ becomes macroscopic compared to the local current $\hat{\cal J}_{b ,x}$. In addition to this, we note that the projection ${\cal Q} $ {\it physically} eliminates the hydrodynamic mode in the currents. Having these in mind, we make one assumption: the Markovian approximation for the term $K_{b x,b ' x' }(u,u' ; t-s)$, meaning that this term rapidly decays in time. 
Under this assumption, we consider the dynamics of {\it macroscopic variable} $\hat{u}$ during the decaying time. In general, macroscopic variables are robust against the short-time evolution; i.e., their values do not change much in time, while the microscopic variables rapidly change in time. Applying this general property to the variables $\hat{u}$ and ${\cal Q}\hat{\cal J}_{b ,x}$, one expects that the time evolution of the variable $\hat{u}$ does not change much for the short-decay time while ${\cal Q}\hat{\cal J}_{b ,x}$ rapidly decays. Therefore, it is {\it physically} reasonable to approximate as 
$\langle ({\cal Q} \hat{\cal J}_{b ,x}) (e^{(t-s) {\cal Q }{\mathbb L} }  \delta (\hat{u} - u') {\cal Q} \hat{\cal J}_{b ' , x'}\rangle_{\rm LM}^u\sim \langle ({\cal Q} \hat{\cal J}_{b , x}) (e^{(t-s) {\cal Q}{\mathbb L} } {\cal Q} \hat{\cal J}_{b ', x'} ) \delta (\hat{u} - u' ))\rangle_{\rm LM}^u =\langle ({\cal Q} \hat{\cal J}_{b , x}) (e^{(t-s) {\cal Q}{\mathbb L} } {\cal Q} \hat{\cal J}_{b ', x'} ) )\rangle_{\rm LM}^u \delta (u - u' )$.
In this physical picture, we proceed one-step further, rewriting the master equation as follows:
\begin{align}
{\partial_t f_t (u) } &= \sum_{a , x} {\delta \over \delta u_{a , x}} (\langle \nabla_x \hat{\cal J}_{a ,x}\rangle_{{\rm LM}}^u f_t (u) ) \nonumber 
\\
&+ \sum_{a ,x}\sum_{a ' , x'}  {\delta \over \delta u_{a ,x}}  \Omega (u)
(\nabla_x \nabla_{x'} K_{a x,a ' x'} (u ) ) {\delta \over \delta u_{a ', x'}}   (f_t (u) / \Omega (u ))  \, , ~~~~\label{zwanzig-3:suppl}\\
  K_{a x,a ' x'}(u)           &:= \int_{0}^{\infty} ds \, 
                           \langle ({\cal Q} \hat{\cal J}_{a ,x}) (e^{s {\cal Q}{\mathbb L} } {\cal Q} \hat{\cal J}_{a ', x'}) \rangle_{\rm LM}^u
                           \, . \label{firstappr0:suppl}
\end{align}
Here, we confine ourselves to consider the near-equilibrium regime in order that we can reasonably replace the average $\langle ...\rangle_{\rm LM}^u$ in Eq.(\ref{firstappr0:suppl}) by an average over the equilibrium distribution that is determined through the total length, pressure and the total energy for a given initial state. For a later convenience in deriving the corresponding Langevin dynamics, we introduce the symmetric and the anti-symmetric coefficients as
\begin{align}
  \begin{split}
  K_{a x,a ' x'}  &= \int_{0}^{\infty} ds \, 
\langle ({\cal Q} \hat{\cal J}_{a ,x}) (e^{s {\cal Q}{\mathbb L} } {\cal Q} \hat{\cal J}_{a ',x'}) \rangle_{\rm eq} \, , \label{firstappr:suppl} \\
  K_{a x,a' x'}^{\rm (S,A)} &:={(1/2)} \left( K_{a x,a ' x'} \pm K_{a ' x' ,a x } \right) \, ,
  % \\  K_{ax,by}^{\rm (A)} &:={1\over 2} \left( K_{ax,by} - K_{by,ax } \right) \, .
  \end{split}
\end{align}
where $\langle ...\rangle_{\rm eq}$ is the average of the the equilibrium distribution. We then rewrite the master equation as
\begin{align}
  {\partial_t f_t (u) } &= \sum_{a , x} {\delta \over \delta u_{a , x}}
                                         \left[  ( \langle \nabla_x  \hat{\cal J}_{a ,x}\rangle_{{\rm LM}}^u
+\sum_{a ' , x'} (\nabla_{x} K_{a x,a ' x'}^{\rm (A)} ) (\nabla_{x'} \lambda_{a ', x'}) )
                                         f_t (u) \right] \nonumber 
\\
&+ \sum_{a, x}\sum_{a ', x'}  {\delta \over \delta u_{a ,x}}  \Omega (u)
     (\nabla_x \nabla_{x'} K_{a x,a ' x'}^{\rm (S)} ) {\delta \over \delta u_{a ' , x'}}   (f_t (u) / \Omega (u ))  \, , ~~~~\label{zwanzig-2:suppl}\\
  \lambda_{a ', x'} &:={\delta \over \delta u_{a ' , x'}} \log \Omega (u) \, . \label{ftth2:suppl2} 
\end{align}

\begin{figure}[t]
  \centering
  \includegraphics[width=100mm]{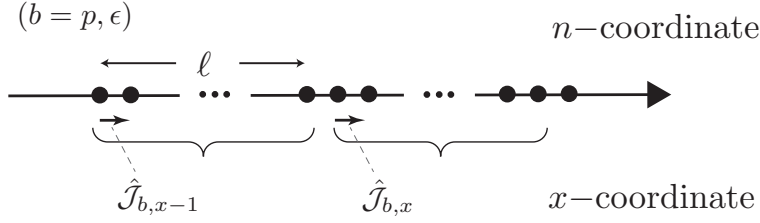}%{./suppl_fig2.pdf} 
  \caption{Schematic indicating that the current is located at the local sites in each sector}
  \label{fig2:suppl}
\end{figure}

\section{Fluctuating hydrodynamics}
We start with the Langevin equation near equilibrium of the form:
\begin{align}
  \partial_t u_{a , x} &=
                     - \nabla_x 
\left[ 
                   \langle \hat{\cal J}_{a , x} \rangle_{{\rm LM}}^u
                   + \sum_{a' , x'} K_{a x,a ' x'} \nabla_{x'} (\delta S/\delta u_{a ' x'}) + \xi_{a , x} (t) \right]  \nonumber \\
      &=               - \nabla_x 
\left[ 
                   \langle \hat{\cal J}_{a , x} \rangle_{{\rm LM}}^u
+\sum_{a ' , x'} K_{a x, a ' x'}^{\rm (A)}   \nabla_{x'} (\delta S/\delta u_{a ' , x'})
  + \sum_{a ' , x'} K_{a x, a ' x'}^{\rm (S)} \nabla_{x'} (\delta S/\delta u_{a ' , x'})
  + \xi_{a , x} (t) \right] \, , \label{L1:suppl}\\
  S &= \log \Omega (u)~ ,
\end{align}
where $S$ is the thermodynamic entropy and we impose the fluctuation-dissipation relation for the noise terms
\begin{align}
  \langle \langle \xi_{a ,x} (t) \xi_{a ' , x'} (t')\rangle \rangle & = 2 K_{a x,a x'}^{\rm (S)} \delta_{a,a'}\delta (t-t') \, ,
\end{align}
where $\langle\langle ... \rangle\rangle$ is a noise average. Here, we should note that the fluctuation-dissipation relation is imposed for the symmetric part only. In addition, we impose the following thermodynamic relation, from the analogy of the standard thermodynamic relation, such as the relation between the entropy, energy and inverse temperature: 
\begin{align}
   {(\delta S / \delta u_{a ', x'})} &= \lambda_{a ', x'} \, . \label{th2}
\end{align}
Through straightforward calculations, one gets the corresponding Fokker-Planck equation for the distribution of $\{ u_{a , x}\}$, denoted by $P_t(u)$, as:
\begin{align}
    {\partial_t P_t (u) } &= \sum_{a , x} {\delta \over \delta u_{a ,x}}
                                         \left[  ( \langle \nabla_x  \hat{\cal J}_{a ,x}\rangle_{{\rm LM}}^u
+\sum_{a ' ,x '} (\nabla_{x} K_{a x,a ' x'}^{\rm (A)} ) (\nabla_{x'} \lambda_{a ' , x'}) )
                                         P_t (u) \right] \nonumber
\\
&+ \sum_{a ,x }\sum_{a ' , x'}  {\delta \over \delta u_{a , x}}  \Omega (u)
     (\nabla_x \nabla_{x'} K_{a x,a ' x'}^{\rm (S)} ) {\delta \over \delta u_{a ' , x'}}   (P_t (u) / \Omega (u ))  \, , 
\end{align} 
which is identical to Eq.(\ref{zwanzig-2:suppl}).

Furthermore, one can proceed on the Langevin-type equation (\ref{L1:suppl}) as follows
\begin{align}
  \partial_t u_{a , x}&=
                     - \nabla_x 
                     \left[  \langle \hat{\cal J}_{a , x} \rangle_{{\rm LM}}^u +
\sum_{a ' ,x'} K_{a x, a ' x'}^{\rm (A)}  \nabla_{x'} \lambda_{a ' , x'} 
                             + \sum_{a ' , x'} K_{a x, a ' x'}^{\rm (S)}  \nabla_{x'}
                             \lambda_{a ' , x'}  + \xi_{a , x} (t) \right] \, \nonumber  \\
                           &\sim
                     - \nabla_x 
                     \left[  \langle \hat{\cal J}_{a , x} \rangle_{{\rm LM}}^u +
\sum_{a '} K_{a, a ' }^{\rm (A)}  \nabla_{x} \lambda_{a ' , x} 
+ \sum_{a ' } K_{a , a ' }^{\rm (S)} \nabla_{x}
                             \lambda_{a ' , x}  + \xi_{a , x} (t) \right]
                             \,  \nonumber \\
  &=
                     - \nabla_x 
                     \left[  \langle \hat{\cal J}_{a , x} \rangle_{{\rm LM}}^u -
\sum_{a ' ,a ''} K_{a, a '' }^{\rm (A)}  \Lambda_{a '',a '}  \nabla_{x} u_{a ' , x} 
    -
\sum_{a ' , a '' } K_{a , a '' }^{\rm (S)}  \Lambda_{a '',a '}  \nabla_{x}
                             u_{a ' , x}  + \xi_{a , x} (t) \right] \, , \label{fht_pre:suppl}
  \\ 
\Lambda_{a '' ,a '}  &:= -({\partial \lambda_{a '' x} / \partial u_{a '  x}})_{\rm eq} \, ,  \label{lambdadef:suppl}
\end{align}
where we replace $ \nabla_{x '}  \lambda_{a ' , x'} $ by $\nabla_x  \lambda_{a ' , x}$ assuming a fast decay in the coefficients $K_{a x , a ' x'}^{\rm (S,A)}$ with respect to the distance $|x-x'|$. This approximation is quite reasonable because we are taking the coarse-graining picture where even one site in the $x$-coordinate already includes $\ell$ sites in the original $n$-coordinate. The coefficients $K^{\rm (S,A)}_{a, a '}$ are defined as $K^{\rm (S,A)}_{a, a '}:= \sum_{x'}K^{\rm (S,A)}_{a x, a ' x'}$. Note that the dependence of $x$ on the coefficients disappears owing to the translational invariance of the system. 
If we further define $D_{a, a '}^{\rm (S,A)} := \sum_{a ''} K_{a , a '' }^{\rm (S,A)} \Lambda_{a '', a '} $, then the hydrodynamics can be written in a more familiar form: 
\begin{align}
  \begin{split}
\partial_t u_{a , x}&=- \nabla_x 
\left[ 
                   \langle \hat{\cal J}_{a , x} \rangle_{{\rm LM}}^u  - \sum_{a '} D_{a, a '}^{\rm (A)} \nabla_x  u_{a ', x}
                   - \sum_{a '} D_{a, a '}^{\rm (S)} \nabla_x  u_{a ' , x}
                           + \xi_{a ,x} (t) \right] \, , \\
               \langle\langle \xi_{a, x} \xi_{a ' , x'}\rangle\rangle &= 2 K_{a,a}^{\rm (S)} \delta_{a,a'}\delta_{x, x'} \delta (t-t') \, . \label{familiar}
                         \end{split} 
\end{align}
Note that the terms $ \langle \hat{\cal J}_{a ,x} \rangle_{{\rm LM}}  - \sum_{a '} D_{a, a '}^{\rm (A)} \nabla_x  u_{a ', x}$ are regarded as a reversible part (or local equilibrium part) in the current, while the terms $- \sum_{a '} D_{a,a '}^{\rm (S)} \nabla_x  u_{a ' , x}$ are the dissipation part that connects to thermal noises $\xi_{a , x} (t)$ via the fluctuation-dissipation relation. 

Using the translational invariance in the system, one can write the Green-Kubo like formula for the bare transport coefficients as
\begin{align}
    K_{a , a'}^{({\rm S,A})}  &= (1/2)  (K_{a , a'} \pm K_{a ' , a } ) \, , \\
                                          K_{a , a'}
                                          = \sum_{x'} K_{a x, a ' x'}
          &=      \int_{0}^{\infty} ds \, 
            \langle ( {\cal Q}  \hat{\cal J }_{a, x}) ( e^{ {\cal Q} {\mathbb  L} s} {\cal Q}  \sum_{x'}\hat{\cal J}_{a ' , x' } ) \rangle_{\rm eq} \,
  \nonumber \\      
          & =(\ell/N) \int_{0}^{\infty} ds \,
            \langle (\sum_x {\cal Q} \hat{ \cal J }_{a ,  x})  ( e^{{\cal Q} {\mathbb L}s} {\cal Q} \sum_{x'} \hat{\cal J}_{a ' , x'}  ) \rangle_{\rm eq}  \\
                                        &\sim (\ell/N) \int_{0}^{\infty} ds \,     \langle (\sum_x {\cal Q} \hat{ \cal J }_{a , x})  ( e^{{\mathbb L}s} {\cal Q} \sum_{x'} \hat{\cal J}_{a ' , x'}  ) \rangle_{\rm eq}
                                          \label{kubo-like:suppl}        \, ,
\end{align}
where we replace $e^{{\cal Q}{\mathbb L} s}$ by $e^{{\mathbb L} s}$, which is addressed again in the subsequent section. Note that ${\cal Q} {\cal J}_{r,x}=(1-{\cal P}) {\cal J}_{r,x}= 0 $, and hence $K_{r,a'}=K_{a,r}=0$ for any $a'$ and $a$. Hence, the above Green-Kubo like formula is used for the other elements.

\section{Ensemble equivalence between local Gibbs and microcanonical distribution}
\subsection{Large Deviation argument and local Gibbs distribution}
We start with a general large deviation theory. We define $\hat{U}_{a,x} = \ell \hat{u}_{a,x}$, which is originally defined in (\ref{cg:suppl}). We also use $U_{a,x}$ and $u_{a,x}$ as c-number values, where $U_{a,x}=\ell u_{a,x}$. We introduce the moment generating function:
\begin{align}
  Z(\xi )  &= \int d\Gamma \, e^{ \sum_{a,x} \xi_{a,x} \hat{U}_{a,x} (\Gamma) }
\end{align}
The phase volume $\Omega (u)$ defined in (\ref{phase-suppl}) is computed via the inverse laplace transform:
\begin{align}
  \Omega (u) &= \Bigl[ \prod_{a' , x'} \int {d\xi_{a', x'} \over 2 \pi i} \Bigr] e^{ -\sum_{a,x} \xi_{a,x} U_{a,x} } Z(\{ \xi \}) \, ,
\end{align}
where the contour is chosen on the complex plane, so that all relevant poles are picked up.
Note that these function should take the following large deviation form for large $\ell$
\begin{align}
  \begin{split}
  Z( \xi ) &\sim e^{\ell \mu (\{ \xi \})} \, , \\
  \Omega (u) &\sim e^{\ell  s ( \{ u \})} \, .
  \end{split}
\end{align}
The large deviation functions connect with each other through the Legendre transform:
\begin{align}
  s( u ) &= \inf_{\{ \xi \}} \left[ \mu ( \xi )       - \sum_{a,x} \xi_{a,x} u_{a,x} \right] \, ,
  \label{lg1:suppl2}\\
  \mu ( \xi ) &= \sup_{ \{u \}} \left[ s ( u  )       + \sum_{a,x} \xi_{a,x} u_{a,x} \right] \, .               \label{lg2:suppl2}
\end{align}
From the relation (\ref{lg1:suppl2}), we solve the following equations to obtain the infimum:
\begin{align}
  \partial \mu (\xi )/ \partial \xi_{a,x} - u_{a,x} &= 0 \, . \label{condvari:suppl2}
\end{align}
Let $\xi_{a,n}^{\ast}$ be a solution of this equation. Using the value $\xi_{a,n}^{\ast}$, the function $s$ is written as
\begin{align}
  s( u ) &=  \mu ( \xi^{\ast} )       - \sum_{a,x} \xi_{a,x}^{\ast} u_{a,x}  \, .   \label{su:suppl2}
\end{align}
Let us define
\begin{align}
  S(u) & := \ell s (u) \, , \\
  \lambda^{a,x} & := -\ell \xi_{a,x}^{\ast} \, .
\end{align}
Then we have the following relation
\begin{align}
(\partial S (u)/ \partial u_{a,x} ) &=   \ell ( \partial s (u)/ \partial u_{a,x} )  \nonumber \\
                                    &= - \ell \xi_{a,x}^{\ast} + \ell \sum_{a' , x'} (\partial \xi_{a' , x'}^{\ast} / \partial u_{a,x} )  (  (\partial \mu / \partial \xi_{a' , x'}^{\ast}) - u_{a' , x'} ) = \lambda_{a,x} \, ,
\end{align}
where we use the conditions (\ref{condvari:suppl2}). This relation corresponds to (\ref{ftth2:suppl2}) and (\ref{th2}).

Having established the formal structure, we next consider the effective Gibbs ensemble, so that the entropy function $S (u) $ is smoothly given via the large deviation structure. Following this criterion, we take the following local Gibbs distribution
\begin{align}
  \hat{\rho}_{\rm LG}  &= e^{ \sum_{a,x} \xi_{a,x}^{\ast} \hat{U}_{a,x} (\Gamma) } /Z_{\rm LG} =
                        e^{ -\sum_{a,x} \lambda_{a,x} \hat{u}_{a,x} (\Gamma) }/Z_{\rm LG} \, ,  \label{llg:suppl2}
\end{align}
where $Z_{\rm LG}$ is the partition function. Note that the variable $\lambda_{a,n}$ is given by the conditions (\ref{condvari:suppl2}), which are essentially equivalent to solving the following solutions for large $\ell$ where the large deviation functions dominate the functional structure:
\begin{align}
  u_{a,x} &= \int d\Gamma \hat{u}_{a,x} \hat{\rho}_{\rm LG} \, .
            \label{3neq}
\end{align}
In addition, we assume that the hydrodynamic motions for coarse-grained stretch variables are very smooth along the $x$-coordinate and that the gradients are very small. Then we employ approximation on the local Gibbs ensemble concering the stretch terms. To this end, let us write the coarse-grained stretch variable in terms of the microscopic variable and look at the detailed expression of the exponent in the the exponential function in the local Gibbs ensemble (\ref{llg:suppl2}) as follows
\begin{align}
  \hat{u}_{r,x} &= (1/\ell) \sum_{n=(x-1)\ell +1}^{x \ell} \sum_{k=1}^{\ell}\hat{r}_{n+k-1} /\ell  = (1/\ell) \sum_{n=-\ell +1}^{\ell -1} \hat{r}_{x\ell + n}  (1- |n|/\ell)  \, , \label{urx-micro:suppl} \\
  \sum_{x} \lambda_{r,x} \hat{u}_{r,x}
  &= \sum_{x} \lambda_{r,x} (1/\ell) \sum_{n=-\ell +1}^{\ell -1} \hat{r}_{x\ell + n}
    (1- |n|/\ell)  \, \nonumber
  \\
                                        &= \sum_x (1/\ell)\sum_{n=1}^{\ell } \hat{r}_{(x-1)\ell + n } \left[ \lambda_{r,x} - (1- n/\ell) (\lambda_{r,x-1} - \lambda_{r,x})\right] \nonumber \\
  &=  \sum_x (1/\ell  )\sum_{n=1}^{\ell} \hat{r}_{(x-1)\ell + n } \lambda_{r,x} + O( \nabla \lambda_r ) \, . \label{exponent-r:suppl}
\end{align}
Note the parameter $\lambda_{r,x}$ is a function of $\{ u_{r,x}\}$. When the hydrodynamic motion is very smooth with small gradients, which is the case in our situation, and hence we drop the gradient contribution in (\ref{exponent-r:suppl}) to employ the following approximation on the local Gibbs expression:
\begin{align}
  \hat{\rho}_{\rm LG} &\sim  \hat{\tilde{\rho}}_{\rm LG}  := \prod_x \hat{\tilde{\rho}}_{\rm LG}^{(x)}  \,  , \label{applg1:suppl} \\
  \hat{\tilde{\rho}}_{\rm LG}^{(x)}  &=e^{ -  \sum_{a} \lambda_{a,x} \hat{\tilde{u}}_{a,x}   } / Z_x \, , \\
  \hat{\tilde{u}}_{a,x}   &:=
                             (1/\ell)\sum_{n=1}^{\ell -1} \hat{c}_{a,(x-1)\ell +n} \, , ~~~a=r,p,\epsilon \, ,
\end{align}
where $Z_{x}$ is the normalization. Note that $\hat{\tilde{u}}_{p,x}= \hat{u}_{p,x}$ and $\hat{\tilde{u}}_{\epsilon,x} = \hat{u}_{\epsilon ,x}$, and hence only stretch part $\hat{\tilde{u}}_{r,x}$ is modified from the original expression $\hat{u}_{\epsilon ,x}$. The parameter $\lambda_{a,x}\,(a=r,p, \,{\rm and} \, \epsilon)$ is determined through
\begin{align}
  \begin{split}
    \langle \hat{\tilde{u}}_{a,x} \rangle_{\rm LG} &= u_{a,x} \, , \\
    ~~{\rm Equivalently,}~~\langle \hat{c}_{a,(x-1)\ell + n } \rangle_{\rm LG} &= u_{a,x}\, , ~~~~n=1,\cdots , \ell -1 \, , ~~a=r,p,\epsilon \, , 
\end{split}
    \label{applg2:suppl}
\end{align}
where $\langle ... \rangle_{\rm LG}$ is the average over the Local Gibbs ensemble $\hat{\tilde{\rho}}_{\rm LG}$ defined in (\ref{applg1:suppl}). The condition (\ref{applg2:suppl}) is equivalent to the condition (\ref{3neq}) for the stretch term also within the approximation neglecting the order of gradient contribution. That is, inserting the expression (\ref{applg2:suppl}) to the expression (\ref{urx-micro:suppl}) yields $\langle \hat{u}_{r,x} \rangle_{\rm LG} = u_{r,x} +(1/2)(u_{r,x+1} - u_{r,x}) \sim u_{r,x}$. From the argument above, we can reasonably employ the local Gibbs ensemble $\hat{\tilde{\rho}}_{\rm LG}$ defined in (\ref{applg1:suppl}) with the condition (\ref{applg2:suppl}) for the asymptotic expression of microcanonical average for momentum and energy currents as discussed below. More concretely, using $\hat{\tilde{\rho}}_{\rm LG}$ is justified as long as we consider the leading-zeroth order contribution of $\{ u_{a,x} \}$ in local equilibrium currents, as we discuss in the subsequent subsection.

\subsection{Asymptotic expression of microcanonical average and projection for momentum and energy currents}
We first note that the coarse-grained stretch current is directly connected to the coarse-grained momentum as in (\ref{currnetcg:suppl}), and so one can exactly write the local microcanonical average as
\begin{align}
  \langle \nabla_x   \hat{\cal J}_{r , x}\rangle_{{\rm LM}}^u & = \nabla_x (-u_{p,x}) \, .
\end{align}  
We next consider the momentum and energy currents and the averages over the local microcanonical distribution. We write them in terms of the current in the original $n$-coordinate as
\begin{align}
 \langle \nabla_x \hat{\cal J}_{b , x}\rangle_{{\rm LM}}^u
  & =  (1/\ell) (\, \langle \hat{j }_{b , x\ell +1} \rangle_{{\rm LM}}^u  - \langle \hat{j }_{b , (x-1)\ell +1} \rangle_{{\rm LM}}^u \, ) \, ,~~~b=p,\epsilon \, .   \label{oct:suppl}
\end{align}
This implies that the average is calculated only for the local site within the $\ell$ sites; see Fig.\ref{fig2:suppl} again. Let us focus on the second term in (\ref{oct:suppl}) and express it more explicitly as:
\begin{align}
  \langle \hat{j }_{b , (x-1) \ell +1}  \rangle_{{\rm LM}}
  &=\int d\Gamma  \, \hat{j }_{b , n }  (\Gamma ) \hat{\rho}_{\rm LM} (\Gamma ) \, \bigr|_{n=(x-1) \ell +1} \nonumber \\
  &=\int d\Gamma \,  \hat{j }_{b , n} (\Gamma ) \left[ \prod_{x'}\prod_{a ' =r,p,\epsilon}\delta (\hat{u}_{a ' ,x'} - u_{a ' , x'})  /\Omega (u )  \right] \,  \Bigr|_{n=(x-1) \ell +1} \, , ~~b=p,\epsilon .  \label{eq1:suppl}
\end{align}
The explicit description (\ref{eq1:suppl}) shows that the local observable is averaged with the microcanonical ensemble for $\ell$ particles.
Resorting to the ensemble equivalence for sufficiently large $\ell$, one can accurately replace the microcanonical ensemble by the local Gibbs ensemble: 
\begin{align}
  \hat{\rho}_{\rm LM} &\cong \hat{\tilde{\rho}}_{\rm LG} \, ,
%\\
%  \hat{\rho}_{\rm LG}  & := \prod_x \hat{\rho}_{\rm LG}^{(x)} \, ,  ~~~\hat{\rho}_{\rm LG}^{(x)}= e^{-\sum_{a } \lambda^{a ,x} \hat{u}_{a , x} } / Z_x \, , 
\end{align}
where $\hat{\tilde{\rho}}_{\rm LG}$ is defined in (\ref{applg1:suppl}).
%The parameter $\lambda^{a , x}$, for all $a$ and $x$, is determined by the conditions:
%\begin{align}
%  \begin{split}
%  \langle \hat{u}_{a , x} \rangle_{\rm LG}^u &= u_{a ,x } \, , ~~~~~ {\rm and~~equivalently} \,    \label{cond2:suppl} \\
%  \langle \hat{c}_{a , n} \rangle_{\rm LG}^u &= u_{a ,x } \, , ~~~~~{\rm for}~~~(x-1)\ell + 1 \le n \le x \ell \, .
%  \end{split}
%\end{align}
%where $\langle ...\rangle_{\rm LG} $ implies the average over the local Gibbs ensemble.
Using the thermal expansion [1], one can expand the current expression up to the second order as follows:
\begin{align}
  \langle  \hat{\cal J}_{b, x}  \rangle_{{\rm LM}}^u  &\cong  \langle  \hat{\cal J}_{b, x}  \rangle_{{\rm LG}} \nonumber  \\
                                         &= \sum_{a '=r,p,\epsilon}A_{b,a'} \delta {u}_{a ' , x}  + \sum_{a,a'=r,p,\epsilon}(1/2) H^{b}_{a,a'}   \delta {u}_{a, x} \delta {u}_{a' x} + O((\delta u)^3) \, ,  ~~b=p,\epsilon \label{nonlin:suppl}
\end{align}
where $\delta {u}_{a , x}:= {u}_{a , x} - {u}^{\rm eq}_{a , x}$ and ${u}^{\rm eq}_{a ,x}$ is an equilibrium value.
The detailed expressions of tensors ${\bm A}$ and ${\bm H}$ are given in Ref.[2].
The ensemble equivalence justifies the recipe that has been employed to obtain the local equilibrium current in the nonlinear fluctuating hydrodynamics in Ref.[2].

\vspace*{0.5cm}
We next consider the projection of the current. Note first the exact relation
\begin{align}
 ( {\cal P}  \hat{{\cal J}}_{r ,x} ) (\Gamma ) & = - \hat{u}_{p,x}(\Gamma ) \, .
\end{align}
Hence, we below focus on $(  {\cal P}  \hat{{\cal J}}_{b ,x} ) (\Gamma )$ for $b=p,\epsilon$. To this end, we note that the mathematical structure is the same as that in (\ref{eq1:suppl}) except that this case contains phase-space dependent variables. From this observation, one can immediately find the connection between the projection to the local Gibbs ensemble. Therefore, with the ensemble equivalence idea, this connection leads to the following approximation: 
\begin{align}
  (  {\cal P}  \hat{{\cal J}}_{b , x} ) (\Gamma )
  &= \int d\Gamma ' \,  \hat{j }_{b , n} (\Gamma ' ) \left[ \prod_{x'} \prod_{a ' =r,p,\epsilon}\delta (\hat{u}_{a ' ,x'} (\Gamma ') - u_{a ', x'} (\Gamma ))  /\Omega (\hat{u}(\Gamma))  \right] \,  \Bigr|_{n=(x-1) \ell +1} \,     
    \nonumber  \\
  &\cong \int d \Gamma ' \, \hat{j }_{b, n} (\Gamma ')  \, { \prod_{x'} e^{ - \sum_{a '=r,p,\epsilon} \hat{\lambda}_{a ' ,x'} (\Gamma ) \,
 \hat{\tilde{u}}_{a ', x'} (\Gamma ')  }
    / \hat{Z} (\Gamma) } \, \bigr|_{n=(x-1)\ell +1} \, , \label{proje1:suppl} ~~~b=p,\epsilon \, , 
\end{align}
where the phase-dependent parameter $\hat{\lambda}_{a , x}(\Gamma)$, for all $a $ and $x$, is determined by the conditions:
\begin{align}
%  \begin{split}
%    \int d\Gamma '   \hat{u}_{a , x} (\Gamma ' ) \, {\prod_{x'} e^{ - \sum_{a '} \hat{\lambda}^{a ' ,x'} (\Gamma ) \hat{u}_{a ', x'} (\Gamma ')} / \hat{Z} (\Gamma) }  &= \hat{u}_{a ,x } (\Gamma ) \, , ~~~{\rm and ~~equivalently , } \\
  \int d\Gamma '   \hat{c}_{a , n} (\Gamma ' ) \, { \prod_{x'} e^{ - \sum_{a '} \hat{\lambda}_{a ' ,x'} (\Gamma )  \hat{\tilde{u}}_{a ', x'} (\Gamma ')
%  \hat{u}_{a ', x '} (\Gamma ')
  } / \hat{Z} (\Gamma) }  &= \hat{\tilde{u}}_{a ,x } (\Gamma ) \,  ,    ~~{\rm for}~~~(x-1)\ell + 1 \le n \le x \ell \, . 
%    \end{split}
\end{align}
From this mathematical structure, one can formally expand this with the same coefficient as in (\ref{nonlin:suppl}):
\begin{align}
  (  {\cal P}  \hat{{\cal J}}_{b , x} ) (\Gamma )  & \cong
                                                     \sum_{a '=r,p,\epsilon}
                                                     A_{b, a '} \delta \hat{\tilde{u}}_{a ' , x}  + \sum_{a,a'=r,p,\epsilon}(1/2) H^{b}_{a,a'}   \delta \hat{\tilde{u}}_{a, x} \delta \hat{\tilde{u}}_{a', x} + O((\delta \hat{\tilde{u}} )^3) \, ,  ~~~b=p,\epsilon . \label{gibbspro:suppl} 
\end{align}

\section{Physical argument on (\ref{kubo-like:suppl}) }
In the previous section, we employed the ensemble equivalence from the physical point of view. With this picture, we revisit the Green-Kubo like formula (\ref{kubo-like:suppl}), focusing on the replacement $e^{{\cal Q}{\mathbb L} s}$ by $e^{{\mathbb L} s}$. Note that the Green-Kubo like formula is used to compute $K_{b,b'}$ with $b,b'=p,\epsilon$.

In order to physically argue on this replacement, we begin by looking at the quantity $\hat{R}_s =e^{s {\cal Q} {\mathbb L}}  ( \sum_{x'} {\cal Q} \hat{\cal J}_{b ' , x'})~(b'=p,\epsilon)$, which obeys the dynamics:
\begin{align}
  \partial_s \hat{R}_s & = ( {\mathbb L} - {\cal P} {\mathbb L }  ) \hat{R}_s \, .
\end{align}
Furthermore, we consider the projection part ${\cal P} {\mathbb L }   \hat{R}_s $ which is given as
\begin{align}
  ( {\cal P} {\mathbb L} \hat{R}_s) (\Gamma ) &= \int d\Gamma ' \left\{ \hat{H}(\Gamma ') , \hat{R}_s (\Gamma ') \right\}
         { e^{ - \sum_{a ' , x'} \hat{\lambda}_{a ' , x'} (\Gamma ) \hat{\tilde{u}}_{a ' , x'} (\Gamma ')} / \hat{Z} ( \Gamma ) }   \nonumber \\
  &= \int d\Gamma ' \left\{
    { e^{ - \sum_{a ' , x'} \hat{\lambda}_{a ' , x'} (\Gamma ) \hat{\tilde{u}}_{a ' , x'} (\Gamma ')} / \hat{Z} ( \Gamma ) }  ,  \hat{H}(\Gamma ')  \right\}  \hat{R}_s (\Gamma ')  \, ,  \label{sa1:suppl}
\end{align}
where $\{ ... , ...\}$ implies the Poisson bracket. For this expression, we expand the local Gibbs distribution from the equilibrium distribution $\hat{\rho}_{\rm eq} (\Gamma ')$. In detail, let $\delta\hat{\tilde{u}}_{a , x} (\Gamma )$ be a deviation from the equilibrium value 
${u}_{a , x}^{\rm eq} $, $\delta\hat{\tilde{u}}_{a , x} (\Gamma ):= \hat{\tilde{u}}_{a , x}(\Gamma ) -{u}_{a , x}^{\rm eq} $. Then, we obtain the following expansion
\begin{align}
  (\ref{sa1:suppl})          &=\int d\Gamma '   \left\{      \hat{\rho}_{\rm eq} (\Gamma ')   \left( 1 + \sum_{a ,x} \delta\hat{\tilde{u}}_{a ,x} (\Gamma ') \sum_{a '} \Lambda_{a , a '}\,  \delta \hat{\tilde{u}}_{a ',x} (\Gamma )           + \cdots \right) ,  \hat{H} (\Gamma ')  \right\} \hat{R}_s (\Gamma ') \, \nonumber  \\
                             &=\int d\Gamma '  \hat{\rho}_{\rm eq} (\Gamma ')   \left( \sum_{a ,x}   \left\{ \delta\hat{\tilde{u}}_{a ,x} (\Gamma ')  ,  \hat{H} (\Gamma ')  \right\} \sum_{a '} \Lambda_{a , a '}\,  \delta \hat{\tilde{u}}_{a ',x} (\Gamma )           + \cdots \right)                               \hat{R}_s (\Gamma ') \, \nonumber \\     
                             &=  -\sum_{a , x}   \left(            \int d\Gamma '        \hat{\rho}_{\rm eq} (\Gamma ')    \hat{R}_s (\Gamma ')   \nabla_x   \hat{\cal J}_{a , x}(\Gamma ')  \right) \sum_{a '} \Lambda_{a , a '}\,  \delta \hat{\tilde{u}}_{a ', x} (\Gamma ) + O( (\delta \hat{\tilde{u}} (\Gamma ))^2) \nonumber \\
                             &=  -\sum_{a , x}   \left(       \nabla_x      \int d\Gamma '        \hat{\rho}_{\rm eq} (\Gamma ')    \hat{R}_s (\Gamma ')         \hat{\cal J}_{a , x}(\Gamma ')  \right) \sum_{a '} \Lambda_{a , a '}\,  \delta \hat{\tilde{u}}_{a ', x} (\Gamma ) + O( (\delta \hat{\tilde{u}} (\Gamma ))^2) \nonumber \\
                              & = O( (\delta \hat{\tilde{u}} (\Gamma ))^2) \, , \end{align}
where $\Lambda_{a,a'}$ is defined in (\ref{lambdadef:suppl}) and the relation (\ref{contmacc:suppl}) is used to write the equation in terms of the current variables. Here, we consider the function $\hat{R}_s$ that satisfies a translational invariance, namely, it does not depend on $x$. From this calculation, one finds that the contribution of the projection part ${\cal P} {\mathbb L} \hat{R}_s$ starts from the second order of $\delta\hat{\tilde{u}}$. This implies $e^{s {\cal Q} {\mathbb L}}$ is dominated by $e^{s {\mathbb L}}$ at near equilibrium assuming that the deviation from the equilibrium value $\delta\hat{\tilde{u}}_{a ,x }$ is small. 
We should also recall that the variable $\hat{\tilde{u}}_{a , x}$ is macroscopic: it does not change much in the short-time evolution (For large $\ell$, one expects that the amplitudes of $(\delta \hat{\tilde{u}} )^2$ should be the order $1/\ell$, and hence the overall structure of the correlation  $\langle (\sum_x {\cal Q}\hat{ \cal J }_{b , x})  ( e^{{\cal Q}{\mathbb L}s} {\cal Q} \sum_{x'} \hat{\cal J}_{b ' , x'}  ) \rangle_{\rm eq}~~(b,b'=p,\epsilon)$ for short time scale is determined by up to the first order.
%We also expect that higher orders can contribute to suppress the hydrodynamic modes in the large time-scale
). From this physical argument, we expect that the replacement by
$\langle (\sum_x {\cal Q}\hat{ \cal J }_{b , x})  ( e^{{\mathbb L}s} {\cal Q} \sum_{x'} \hat{\cal J}_{b ' , x'}  ) \rangle_{\rm eq}$ should not cause a significant error for computing the time-integral, even at quantitative level.

\section{Bare transport coefficients}
\subsection{Matrix structure}
Note that the condition (\ref{applg2:suppl}) is written as
\begin{align}
  \langle \hat{c}_{a , n} \rangle_{\rm LG} =
\int d\Gamma_n \hat{c}_{a , n} e^{-\sum_{a=r,p,\epsilon}\lambda_{a , n} \hat{c}_{a ,n}} / Z_n=
  u_{a ,x } \, ,~~~\lambda_{a,n}=\lambda_{a,x}/\ell \label{abv:suppl}
\end{align}
for $(x-1)\ell + 1 \le n \le x \ell$. Here, $\Gamma_n$ is the phase space for the $n$th particle only and $Z_n$ is the normalization. Based on this expression, we first recall that 
$\Lambda_{a,a'} = -(\partial \lambda_{a , n}  / \partial u_{a ' , x} )_{\rm eq}$ gives the inverse susceptibility matrix.
 Then, one readily find its components from the susceptibility matrix 
  $[ {\bm \Lambda}^{-1} ]_{a ' , a} :=-(\partial u_{a' , x}  / \partial \lambda_{a , n} )_{\rm eq}= \langle \hat{c}_{a' , n}  \hat{c}_{a , n} \rangle_{\rm eq}  - \langle \hat{c}_{a' , n} \rangle_{\rm eq} \langle \hat{c}_{a , n} \rangle_{\rm eq}$.
  Note that $\Lambda_{a ' , a} = -(\partial u_{a' , x}  / \partial \lambda_{a , n} )_{\rm eq}$ is independent of the site and hence, the matrix $\Lambda_{a,a'}$ is also site-independent. In addition, the structure of the inverse susceptibility matrix elements $\Lambda_{a,a'}$ are given by
\begin{align}
 &
\left(
\begin{array}{ccc}
\Lambda_{r, r} & 0 & \Lambda_{r ,\epsilon} \\
0 & \Lambda_{p,p} & 0 \\
\Lambda_{\epsilon , r} & 0 & \Lambda_{\epsilon , \epsilon} \\
\end{array}
  \right) \, ,
  \end{align}
 with the following expressions for the finite matrix elements
  \begin{align}
    \Lambda_{r , r} &= {(1/ {\cal N})}(\langle V (\hat{r}); V (\hat{r}) \rangle_{\rm eq} + \beta^{-2}/2 ) \, , ~~~
\Lambda_{r , \epsilon} = {-(1 / {\cal N} )}  \langle V (\hat{r}); \hat{r} \rangle_{\rm eq} \, ,\\
\Lambda_{p , p} &= \beta \, ,\\
\Lambda_{\epsilon  , r} &= \Lambda_{r , \epsilon} \, , ~~~
\Lambda_{\epsilon , \epsilon} = {(1 / {\cal N})}  \langle \hat{r}; \hat{r} \rangle_{\rm eq} \, , \\
{\cal N}&=  
\langle \hat{r};\hat{r} \rangle_{\rm eq} \langle V(\hat{r}); V(\hat{r}) \rangle_{\rm eq} 
- \langle \hat{r}; V(\hat{r})  \rangle_{\rm eq}^2  
+ {(\beta^{-2} / 2)}\langle \hat{r};\hat{r} \rangle_{\rm eq}  \, ,
\end{align}
where $\beta$ is the inverse temperature and $\langle \hat{a} ; \hat{b} \rangle_{\rm eq}:=\langle \hat{a}  \hat{b} \rangle_{\rm eq}-\langle \hat{a} \rangle_{\rm eq} \langle  \hat{b} \rangle_{\rm eq}$.

Concerning the matrices ${\bm K}$ and ${\bm D}$, the time-reversal symmetry leads to the relations $K_{p, r}= - K_{r,p}$, $K_{\epsilon , p}= - K_{p, \epsilon}$, and $K_{a,a'}=K_{a',a}$ otherwise.  Note, however, that one cannot impose any other constraints on the matrices only from the symmetry argument. Hence, at the formal level, we have the following matrix structure on ${\bm K}^{\rm (S,A)}$:
\begin{align}
{\bm K}^{\rm (S)} &=               \left(
              \begin{array}{ccc}
                0, & 0, & 0 \\
                0, & K_{p,p} , & 0 \\
                0, & 0, &  K_{\epsilon, \epsilon} 
                \end{array}
                          \right) \, , ~~~~~~
{\bm K}^{\rm (A)} =               \left(
              \begin{array}{ccc}
                0, & 0, & 0 \\
                0, & 0 , & K_{p ,\epsilon} \\
                0, & -K_{p ,\epsilon }, &  0 
                \end{array}
                          \right) \, . \label{kmatrix}
\end{align}
Using the matrix elements of ${\bm \Lambda}$ and ${\bm K}$,  we find the following matrix structure on ${\bm D}$:
\begin{align}
  \begin{split}
  {\bm D}^{\rm (S)} &=
              \left(
              \begin{array}{ccc}
                0, & 0, & 0 \\
                0, & K_{pp} \Lambda_{p , p}, & 0 \\
                K_{\epsilon,\epsilon} \Lambda_{ \epsilon ,  r}
                , &  0, &  K_{\epsilon \epsilon} \Lambda_{\epsilon , \epsilon}
                \end{array}
                          \right) \, , ~ \\
                          {\bm D}^{\rm (A)} &= 
              \left(
              \begin{array}{ccc}
                0, & 0, & 0 \\
                K_{p\epsilon} \Lambda_{\epsilon , r}
                , & 0, &  K_{p ,\epsilon}\Lambda_{\epsilon , \epsilon}
                \\
                0, &  -K_{p\epsilon} \Lambda_{p , p}, & 0
                \end{array}
              \right) \, .
              \end{split}
\end{align}
\subsection{Numerical calculations}
We consider the Fermi-Pasta-Ulam-Tsingou (FPUT) model whose potential is given by
\begin{align}
  V(\hat{r}) &= (1/2)\hat{r}^2 +  (K_3/3)\hat{r}^3+  (K_4/3)\hat{r}^3 \, ,
\end{align}
and also consider the Green-Kubo like formula:
\begin{align}
  K_{a , a '} &=\int_{0}^{\infty} dt \, C_{a , a ' } (t) \, , \\
  C_{a , a ' } (t)        &= (\ell/N)     \langle (\sum_x {\cal Q} \hat{ \cal J }_{a , x})  ( e^{{\mathbb L}t} {\cal Q} \sum_{x'} \hat{\cal J}_{a ' , x'}  ) \rangle_{\rm eq}            \, . 
\end{align}
For the projection, we use the ensemble equivalence technique (\ref{gibbspro:suppl}), employing up to the second order with respect to $\delta \hat{u}(\Gamma)$. Furthermore, we confine ourselves to consider the equilibrium distribution characterized by the temperature only. We do not show the results for finite pressure cases here because they give qualitatively the same results. In addition, we fix the parameters $(k_3, k_4,T)=(2.0, 1.0 , 3.0)$ as in the main text. We perform the numerical calculations for the system size $N=2^{15}$. In Fig. \ref{fig4:suppl}, we show the time dependence of the correlation functions for possible combinations of $a$ and $a '$. The function $C_{a , a '}^{(0)} (t)$ is a standard correlation function defined as
\begin{align}
  C_{a , a '}^{(0)} (t)  &= (1/N) \langle (\sum_n \hat{ j }_{a , n})  ( e^{{\mathbb L}t} \sum_{n'} \hat{j}_{a ' , n'}  ) \rangle_{\rm eq}            \, . 
\end{align}
Below, we list properties for each correlation function.
\begin{itemize}
\item $C_{p,p} (t)$: The standard correlation without subtracting structure, $C_{p,p}^{(0)} (t)$ is a constant at $t=\infty$, because the momentum current has a finite overlap with the conserved quantity in the inner product, i.e., $\langle (\sum_n \hat{j}_{p,n} ) (\sum_{n'} \hat{c}_{r,n'} ) \rangle_{\rm eq} \neq 0$. In case of the function $C_{p,p} (t)$, this property is resolved because of the subtracting structure with projection. Although the second order approximation (\ref{gibbspro:suppl}) is not a perfect projection, $C_{pp} (t=\infty ) \to 0 $ is satisfied for sufficiently large coarse-graining length $\ell$. For small $\ell$, there are humps for every time period $\ell/c$. As increasing $\ell$, the amplitudes of humps become smaller, and the overall functional forms are eventually collapsed onto the same curve where only short time scale has the finite value. The collapsed curve is zero for $t\gtrsim 10.0$. From this structure, we can estimate $K_{p,p} \sim 0.2 \times 10$. 
\item $C_{p,\epsilon } (t)$: Note $C_{p, \epsilon } (0)=0$ from the time-reversal symmetry. For the finite times, there is no reason the correlation must be small from the symmetry argument alone. In Fig.\ref{fig4:suppl}, we observe that the overall functional form of $C_{p,\epsilon} (t)$ is collapsed onto the same curve for sufficiently large $\ell$. The collapsed curve is almost zero for $t\gtrsim 5.0$. From this structure, one find $K_{p,\epsilon}\sim 0.2 \times 10^{-2}  $ which means negligibly small within the present numerical accuracy, and hence one can practically regard it as $K_{p,\epsilon}\sim 0.0$.
%Although this is small, one cannot deny that finite values of $K_{p,\epsilon}$ and $K_{\epsilon,p}$ might cause macroscopic effects.
\item $C_{\epsilon, \epsilon } (t)$: Note that $ C_{\epsilon , \epsilon}^{(0)} (t)$ shows a power-law decay, leading to the divergence of the heat conductivity. However, the power-law behavior is suppressed in $C_{\epsilon , \epsilon} (t) $ and the integration is saturated as shown in the main text (Not shown here. See the main text.). Similarly to $C_{p,p} (t)$, there are hump-structure for small $\ell$ for every time period $\ell/c$. As increasing $\ell$, the amplitudes of humps become smaller, and the overall functional forms are eventually collapsed onto the same curve where only short time scale has the finite value. The saturation in the integration needs longer time ($\sim 100.0$) than the above cases. From this structure, we can estimate $K_{\epsilon,\epsilon} \sim 0.1 \times 10$. 
 
\end{itemize}

\begin{figure}[t]%[htbp]
  \begin{center}
    \begin{tabular}{c}
      % 1
      \begin{minipage}{0.34\hsize}
        \begin{center}
                \includegraphics[width= \textwidth]{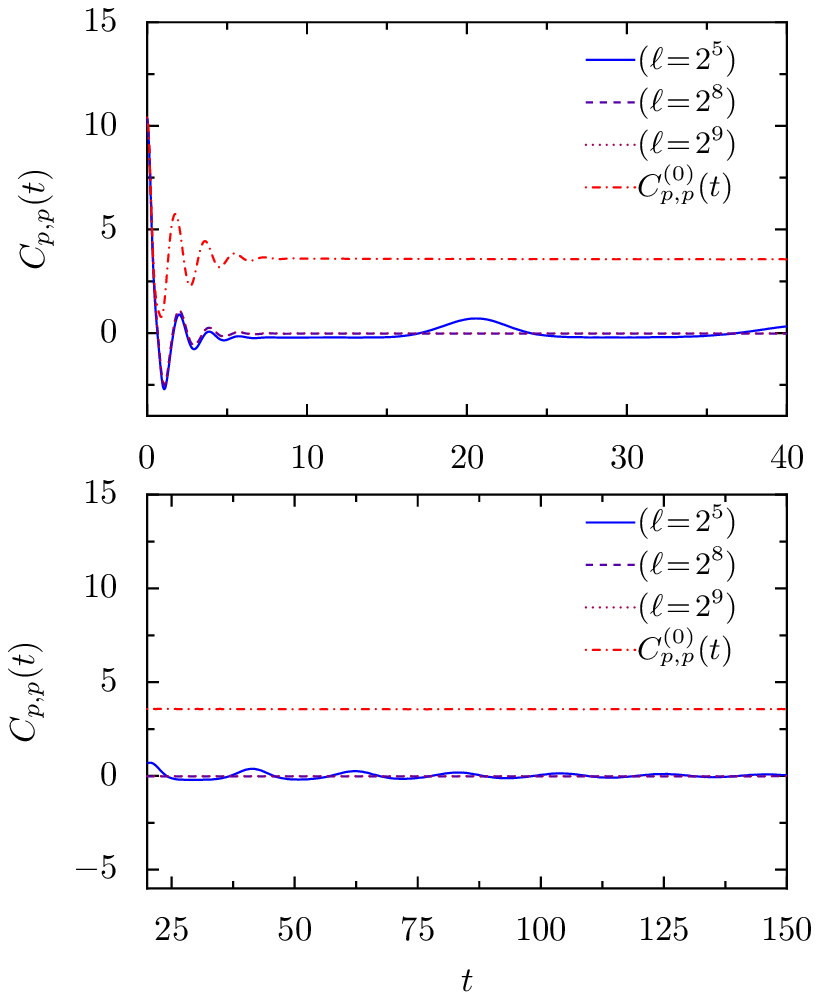}%{./suppl_pp.pdf}
%          \hspace{1.6cm} Case4: $C_{p,p}$.
        \end{center}
      \end{minipage}

      % 2
      \begin{minipage}{0.34\hsize}
        \begin{center}
                \includegraphics[width= \textwidth]{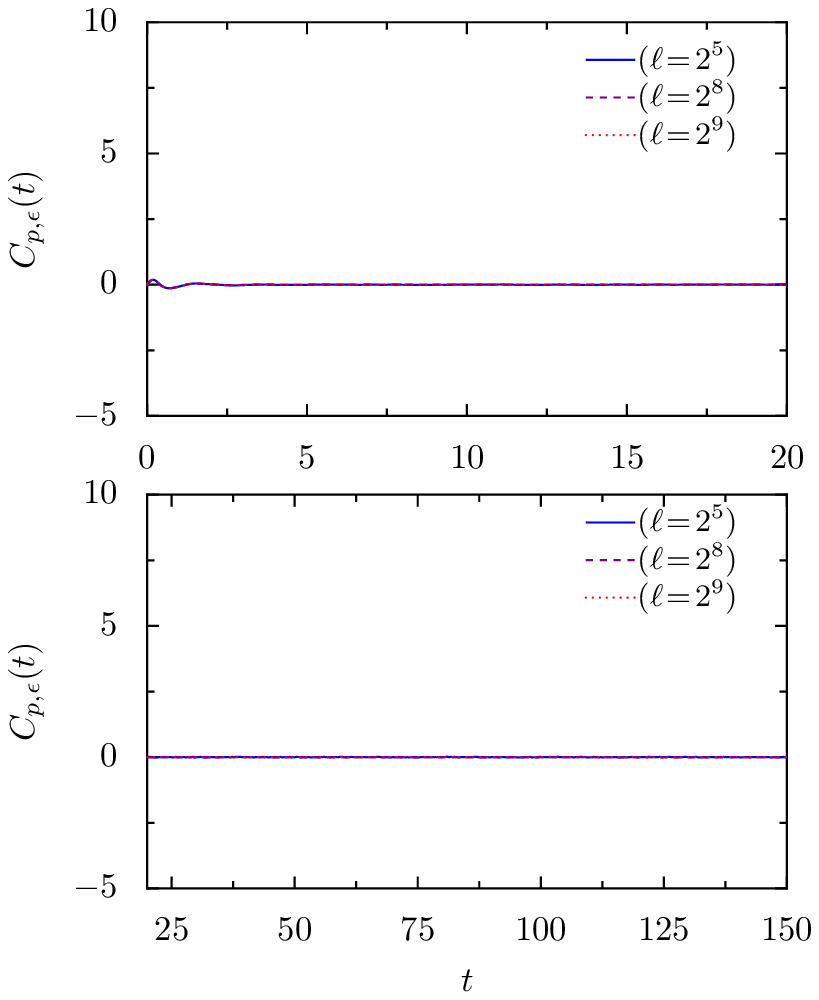}%{./suppl_pe.pdf}
%                \hspace{1.6cm} Case4: $C_{p, \epsilon}$.
        \end{center}
      \end{minipage}
      % 3
      \begin{minipage}{0.34\hsize}
        \begin{center}
                \includegraphics[width= \textwidth]{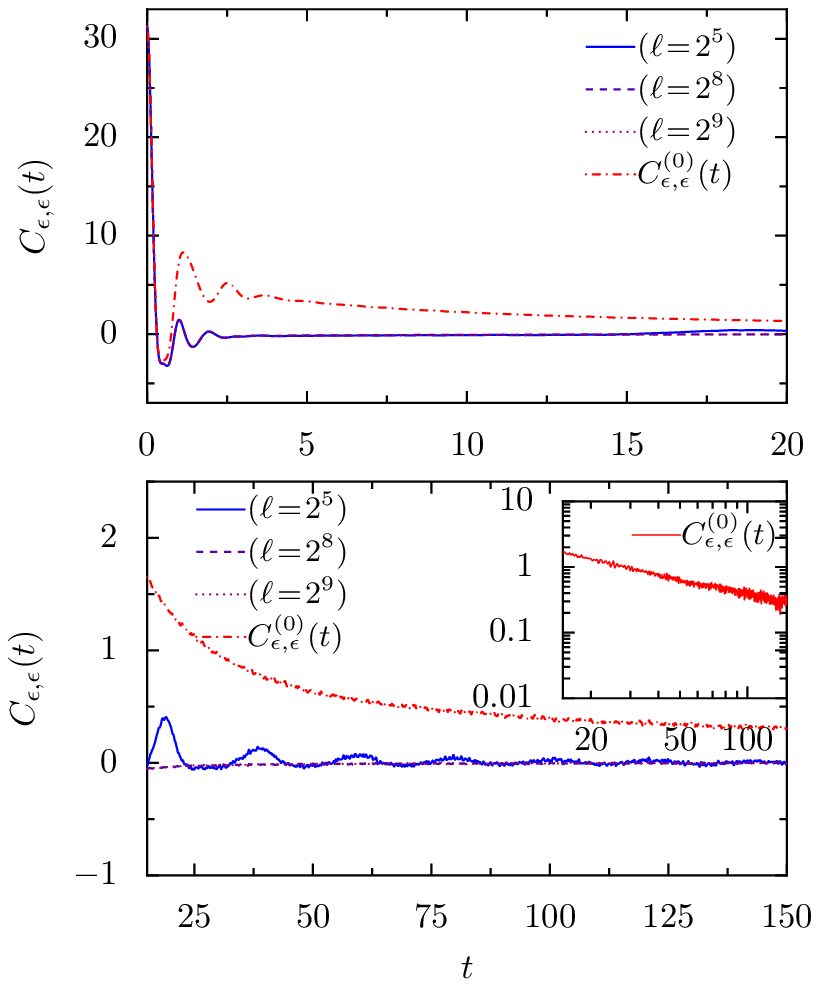}%{./suppl_ee.pdf}
%          \hspace{1.6cm} Case4: $C_{\epsilon, \epsilon}$. 
        \end{center}
      \end{minipage}
    \end{tabular}
    \caption{\label{fig4:suppl}The time-evolution for the correlation functions of $C_{p,p}$, $C_{p,\epsilon}$, and $C_{\epsilon,\epsilon}$. Parameters: $(k_3 , k_4 , T)=(2.0,1.0,3.0)$.}
  \end{center}
\end{figure}
\subsection{Relevant expression of Green-Kubo like formula}
In numerical calculation, we have looked at the dependence of the second order terms in the projection (\ref{gibbspro:suppl}) on the coarse-graining length $\ell$. As we have discussed in Sec.VI, the second order contribution is expected to be smaller as increasing $\ell$, since $(\hat{\delta \tilde{u}}^2)$ should be the order of $1/\ell$. In numerical calculation, we have confirmed this tendency. This suggests that the bare transport coefficient can be obtained only with the first order expansion in (\ref{gibbspro:suppl}) with $\ell \to \infty$ after taking $N\to \infty$ in the Green-Kubo like formula (\ref{kubo-like:suppl}). Hence, the relevant expression for the Green-Kubo like formula is written as
\begin{align}
\begin{split}
 K_{a,a'}&= \lim_{\ell\to\infty}\lim_{N\to\infty}(\ell/N) \int_{0}^{\infty} ds \,   
  \langle (\sum_x {\cal Q} \hat{ \cal J }_{a , x})  ( e^{{\mathbb L}s} {\cal Q} \sum_{x'} \hat{\cal J}_{a ' , x'}  ) \rangle_{\rm eq} \, , \\
  (  {\cal P}  \hat{{\cal J}}_{b , x} ) (\Gamma )  & =
                                                     \sum_{a '=r,p,\epsilon}
                                                     A_{b, a '} \delta \hat{\tilde{u}}_{a ' , x}  \, .
                                                     \end{split}
\end{align}

\section{Thermodynamic structure: Stochastic thermodynamics}
We discuss the thermodynamic structure for the equations
\begin{align}
  \begin{split}
\partial_t u_{a , x}&=- \nabla_x 
\left[ 
  \langle \hat{\cal J}_{a , x} \rangle_{{\rm LG}}^u  + \sum_{a '}
  K_{a , a '}^{({\rm A})} \nabla_x  \lambda_{a ', x}
  + \sum_{a '} K_{a , a '}^{({\rm S})} \nabla_x  \lambda_{a ', x}
                           + \xi_{a ,x} (t) \right] \, , \\
                           \langle \langle \xi_{a ,x} (t)  \xi_{a ' ,x '} (t') \rangle \rangle &= 2 K_{a , a}^{({\rm S} )}  \delta_{x,x'}   \delta (t -t') \, ,
 \end{split}\label{td_fht}
\end{align}
which is equivalent to Eq.(\ref{familiar}). Note that we use the local Gibbs ensemble for the local equilibrium current.
We transform from the variable $u_{a , x}$ to the new variable $h_{a , x}$ defined as
\begin{align}
h_{a ,x} &= \ell \sum_{x' = 0}^{x} u_{a , x'} \, ,
\end{align}
where we set $u_{a , x'=0} =0 $.
The equation (\ref{td_fht}) is then written with this new variable as
\begin{align}
  \partial_t h_{a , x}&=- {\cal J}^{\rm le}_{a,x}
                             - \sum_{a '} K_{a , a '}^{({\rm S})} \nabla_x \lambda_{a ', x}                 + \xi_{a ,x} (t) \, , \label{td_fht_h} \\
 {\cal J}^{\rm le}_{a,x} &:=  \langle \hat{\cal J}_{a , x} \rangle_{{\rm LG}}
                             + \sum_{a '} K_{a , a '}^{({\rm A})} \nabla_x \lambda_{a ', x}            \, ,             
\end{align}
where $\langle \hat{\cal J}_{a , x} \rangle_{{\rm LG}}$ is the representation of $\langle \hat{\cal J}_{a , x} \rangle_{{\rm LG}}^u$ using the variables $\{ h_{a ,x } \}$, and ${\cal J}^{\rm le}$ is a local equilibrium part in the current. The variables $\{ \lambda_{a, x} \}$ are also regarded as functions of $h$.

Let $P_F^h $ be a transition probability of a given forward path $\{ h_{a , x} (t=0)\} \to \cdots \to \{ h_{a , x} (t=\tau) \}$, and let $P_B^h $ be a transition probability for the corresponding backward path $\{ \tilde{h}_{a , x} (t=\tau)\} \to \cdots \to \{ \tilde{h}_{a , x} (t=0) \}$, where $\tilde{\,}$ implies the time-reversal of the variables. From the one-to-one mapping between the variables $h$ and $u$, one can respectively define $P_F^u $ and $P_B^u $ for the forward and backward transition probabilities in terms of the variables $u_{a ,x}$, i.e., $\{ u_{a , x} (t=0)\} \to \cdots \to \{ u_{a , x} (t=\tau) \}$ for the forward path, and $\{ \tilde{u}_{a , x} (t=\tau )\} \to \cdots \to \{ \tilde{u}_{a , x} (t=0) \}$ for the backward path, respectively. Note here the relations:
\begin{align}
  \begin{split}
        \tilde{\partial_t h}_{r,x}
    &= - \partial_t h_{r,x} \, , ~  \tilde{\partial_t h}_{p,x} = \partial_t{h}_{p,x} \, , ~  \tilde{\partial_t h}_{\epsilon ,x} = - {\partial_t h}_{\epsilon ,x}  \, , \\  
    \tilde{\cal J}^{\rm le}_{r,x}
    &= -  {\cal J}^{\rm le}_{r,x} \, , ~~~~~~  \tilde{\cal J}^{\rm le}_{p,x} = {\cal J}^{\rm le}_{p,x} \, , ~~\,~~~  \tilde{\cal J}^{\rm le}_{\epsilon ,x} = {\cal J}^{\rm le}_{\epsilon ,x}  \, , \\  
 \tilde{\lambda}_{r,x}
 &= \lambda_{r,x}\, ,~~~~~~~~~  \tilde{\lambda}_{p,x} = -\lambda_{p,x}\, , ~~~\,\tilde{\lambda}_{\epsilon ,x} = \lambda_{\epsilon,x} \,
 \end{split} \label{tr}
\end{align}                          

From the Gaussian property of the noise term in (\ref{td_fht_h}), one can immediately finds the following relation
\begin{align}
  P_B^u/P_F^u &=  P_B^h/P_F^h  = e^{ -\int_0^{\tau } dt \,Q_t } \, , \\
  Q_t &=  {(1/4)} \sum_{x} \sum_{a,a'}
        ( \tilde{\partial_t h}_{a,x} + \tilde{\cal J}^{\rm le}_{a,x}+ \sum_{a''}K^{({\rm S})}_{a,a''} \nabla_x \tilde{\lambda}_{a'',x}  )
        [({\bm K}^{({\rm S})})^{-1} ]_{a,a'}
        ( \tilde{\partial_t h}_{a',x} + \tilde{\cal J}^{\rm le}_{a',x} +\sum_{a'''}K^{({\rm S})}_{a',a'''} \nabla_x \tilde{\lambda}_{a''',x}  ) \, \nonumber \\
 &- {(1/4)} \sum_{x} \sum_{a,a'}
        ( \partial_t h_{a,x} + {\cal J}^{\rm le}_{a,x}+ \sum_{a''}K^{({\rm S})}_{a,a''} \nabla_x {\lambda}_{a'',x}  )
        [({\bm K}^{({\rm S})})^{-1} ]_{a,a'}
   ( \partial_t h_{a',x} + {\cal J}^{\rm le}_{a',x}+ \sum_{a'''}K^{({\rm S})}_{a',a'''} \nabla_x {\lambda}_{a''',x}  ) \, \nonumber \\
 &=-\sum_{x} \sum_{a} ( \partial_t h_{a,x} + {\cal J}^{\rm le}_{a,x} ) \nabla_x  \lambda_{a,x} \, \nonumber \\
              &=- \sum_{x} \sum_{a}  \partial_t h_{a,x}  \nabla_x  \lambda_{a,x}
=\sum_{x} \sum_{a} \partial_t u_{a,x}  \lambda_{a,x} \, .                
  \end{align}
  where we use the relations (\ref{tr}) and the matrix structure of ${\bm K}^{({\rm A})}$ and ${\bm K}^{({\rm S})}$ in (\ref{kmatrix}). In addition, we note that the local equilibrium current does not contribute to the entropy production. 
From this structure with the standard argument in the stochastic thermodynamics, one can immediately find that the total entropy production $S_{\rm tot}$ during $\tau$ is given as follows
\begin{align}
  S_{\rm tot} &= \int_{0}^{\tau} dt \langle Q_t \rangle_0 -\langle \ln f_{\tau} (\{ \tilde{u} \}) \rangle_0
                +\langle \ln f_{0} (\{ u \}) \rangle_0  \, , 
\end{align}
where $f_{t} (\{ u \})$ is the distribution function at time $t$ and $\langle ... \rangle_{0}$ is an average over the initial distribution. The equation of fluctuating hydrodynamics steadily evolve in time satisfying the nonnegativity of the total entropy production rate $ {\partial_{\tau} S_{\rm tot}} \ge 0 $.

\vspace*{2.0cm}
%\begin{thebibliography}{99}
%\bibitem{spohn2014:suppl}
\noindent
$[1]$ One expands the local Gibbs ensemble from the equilibrium values
$\lambda_{a , {\rm eq}  }$:   
  \begin{align*}
\hspace{10pt}   
 \hat{\tilde{\rho}}_{\rm LG} &=\hat{\rho}_{\rm eq} + (\partial \hat{\tilde{\rho}}_{\rm LG} / \partial \lambda_{a , x} )_{\rm eq} \delta \lambda_{a ,x} + (1/2)(\partial^2 \hat{\tilde{\rho}}_{\rm LG} /  \partial \lambda_{a , x}\partial \lambda_{a ' , x '}
    )_{\rm eq}  \delta \lambda_{a ,x} \delta \lambda_{a' ,x'}   + \cdots ,
  \end{align*}
  where $\delta \lambda_{a ,x }  =  \lambda_{a ,x } - \lambda_{a , {\rm eq}  }$. One also expands the conjugate variables of the c-number quantity $u_{a , x}$ from the equilibrium value $u_{a , {\rm eq} }$ as
  \begin{align*}
 \delta \lambda_{a , x}  &= (\partial \lambda_{a , x} / \partial u_{a ' , x})_{\rm eq} \delta u_{a ' ,x}                              +  (1/2)(\partial^2 \lambda_{a ,x} /  \partial u_{a' , x} \partial u_{a '' , x } )_{\rm eq}  \delta u_{a ' ,x} \delta  u_{a '' ,x}   + \cdots , 
  \end{align*}
  where $\delta u_{a ,x }  =  u_{a ,x } - u_{a , {\rm eq} }$. This combination gives the expansion of the local Gibbs ensemble in terms of $\{ u_{a ,x}\}$. Averaging the currents over this expression yields the local equilibrium current. \\
$[2]$ H. Spohn, {\it Nonlinear Fluctuating Hydrodynamics for Anharmonic Chains}, arXiv:1305.6412, J. Stat. Phys. {\bf 154}, 1191 (2014). \\

\end{widetext}

\end{document}